\documentclass[10pt,twocolumn]{revtex4-1}
\usepackage[utf8]{inputenc}
\usepackage[english]{babel}
\usepackage{float}
\usepackage{graphicx}
\usepackage{bbm}
\usepackage{physics}
\usepackage{xcolor}
\usepackage{mathtools}
\usepackage{multirow}
\usepackage[section]{placeins}
\usepackage{lipsum, babel}
\usepackage{amsmath,amsfonts}
\newcommand{\mean}[1]{\left\langle{#1}\right\rangle}

\begin{document}
	
	\title{Coupling particle-based reaction-diffusion simulations with reservoirs mediated by reaction-diffusion PDEs.}
	\author{Margarita Kostré$^{\dagger}$, Christof Schütte$^{\dagger, \text{\ddag}}$, Frank Noé$^{\text{\ddag},a)}$ and Mauricio J. del Razo$^{\text{\ddag},a)}$}
	
	
	\begin{abstract}
		\textbf{Abstract:} Open biochemical systems of interacting molecules are ubiquitous in life-related processes. However, established computational methodologies, like molecular dynamics, are still mostly constrained to closed systems and timescales too small to be relevant for life processes. Alternatively, particle-based reaction-diffusion models are currently the most accurate and computationally feasible approach at these scales. Their efficiency lies in modeling entire molecules as particles that can diffuse and interact with each other. In this work, we develop modeling and numerical schemes for particle-based reaction-diffusion in an open setting, where the reservoirs are mediated by reaction-diffusion PDEs. We derive two important theoretical results. The first one is the mean-field for open systems of diffusing particles; the second one is the mean-field for a particle-based reaction-diffusion system with second-order reactions. We employ these two results to develop a numerical scheme that consistently couples particle-based reaction-diffusion processes with reaction-diffusion PDEs. This allows modeling open biochemical systems in contact with reservoirs that are time-dependent and spatially inhomogeneous, as in many relevant real-world applications. 
	\end{abstract}
	
	\maketitle 
	
	\noindent $^{\dagger}$ Zuse Institute Berlin, Germany.
	
	\noindent $^{\text{\ddag}}$Freie Universität Berlin, Department of Mathematics and Computer Science, Germany
	
	\noindent $^{a)}$Corresponding authors. E-mails: \\ \texttt{m.delrazo@fu-berlin.de} \ \ \ \ \ \texttt{frank.noe@fu-berlin.de}

	\section{Introduction}
	
	
	Complex systems of interacting particles/agents that exchange energy and matter with a large reservoir are extremely common, from a city exchanging infected citizens at airports during a pandemic to a living cell exchanging chemicals with its environment. In the context of molecular biology, most biochemical reaction systems, either inside living cells or composed by them, interact with some form of reservoir \cite{qian2007phosphorylation}.	They also consume chemical energy for their survival, produce waste and dissipate heat; they operate in an open non-equilibrium setting. In terms of physical chemistry, every living system must be an open system —a closed system has no life \cite{qian2007phosphorylation}. It is thus fundamental to develop models of biochemical reaction systems capable of exchanging materials and energy with their environment. Although this is the guiding motivation for this work, the results here presented can also be applied in other areas such as agent-based modeling.
	
	Molecular dynamics (MD) is theoretically capable of modeling biochemical systems accurately at cellular and sub-cellular scales. However, in timescales relevant to life processes, even a system with one or two molecules is already large enough to render any MD simulation computationally unfeasible. Moreover, although there is on-going research on open MD systems \cite{agarwal2015molecular, delle2019liouville, delle2019molecular}, established computational protocols rely by design on the simulated MD system being closed \cite{delle2019liouville}. Consequently, the most accurate and computationally reliable simulations for open systems at these scales are based on stochastic particle-based reaction-diffusion (PBRD) models. Novel methods are emerging capable of coupling PBRD with MD, integrating the accuracy of MD into efficient PBRD simulations \cite{dibak2018msm, sbailo2017efficient, vijaykumar2015combining}. This, along with other recent research, points out the need for hybrid reaction-diffusion methods that are accurate, open and consistent across multiple scales. 
	
	In this paper, we concentrate on multiscale models for PBRD simulations of open systems, where we focus on coupling particle-based simulations to macroscopic particle/chemical reservoirs. These reservoirs are given by a mean concentration of chemical species that can vary in time and space. We model these reservoirs as reaction-diffusion partial differential equations (PDEs). The goal of this work is to achieve a mathematically consistent coupling between the PBRD simulations and the reaction-diffusion PDEs. This entails two major challenges:
	\begin{itemize}
		\item Determine how to consistently couple a particle-based simulation to a constant concentration chemical reservoir. Note the particle-based simulation will be in the grand canonical ensemble since the number of particles is not constant. This will be solved by calculating the mean-field for an open system of diffusing particles.
		\item Calculate the relations between the reaction rates and diffusion coefficients in the particle-based simulation and the macroscopic reaction-diffusion PDE. This can be especially cumbersome for second-order reactions. We will solve this by calculating the mean-field for reaction-diffusion systems with up to second-order reactions.
	\end{itemize}  
	
	Unlike the case of homogeneous reaction theory \cite{anderson2015stochastic, kurtz1972relationship, qian2011nonlinear}, the connection between microscopic, mesoscopic and macroscopic scales for reaction-diffusion phenomena is still a matter of recent research \cite{arnold1980consistency, feng1996hydrodynamic, hellander2014reaction, isaacson2008relationship, isaacson2013convergent}. One of the main difficulties to establish this connection is to relate the macro and microscopic rates for second-order reactions. We will present  results on the relation between these rates at first order. The results for the most general case are currently a work in progress \cite{kostreHydrolimit}. 
	
	Other very relevant work in this area is \cite{franz2013multiscale}, where the authors introduce a method to couple Brownian dynamics simulations with mean-field PDEs. However, they do not implement their methods for the nontrivial case of second-order reactions, and they conceptually do not consider the PDE domain as a reservoir. The work \cite{smith2018auxiliary} builds and improves on these ideas. The authors include a second-order reaction example. However, the relation between the particle-based and PDE reaction rate does not seem correct. It is based on \cite{erban2009stochastic}, where the diffusion is included in the rate relation, which is only correct if diffusion is averaged out. In this work, we show that the diffusion coefficient does not play a role in the relation between the microscopic and the macroscopic reaction-diffusion PDE reaction rates.
	
	The theory and numerical implementation details are explained in this paper;
	the code is available in GitHub under an MIT license \cite{kostre2020GitHub}.
	
	\section{Reaction-diffusion models}\label{sec:RDM}
	
\begin{figure}[htb]
		\center
		\includegraphics[width=\columnwidth]{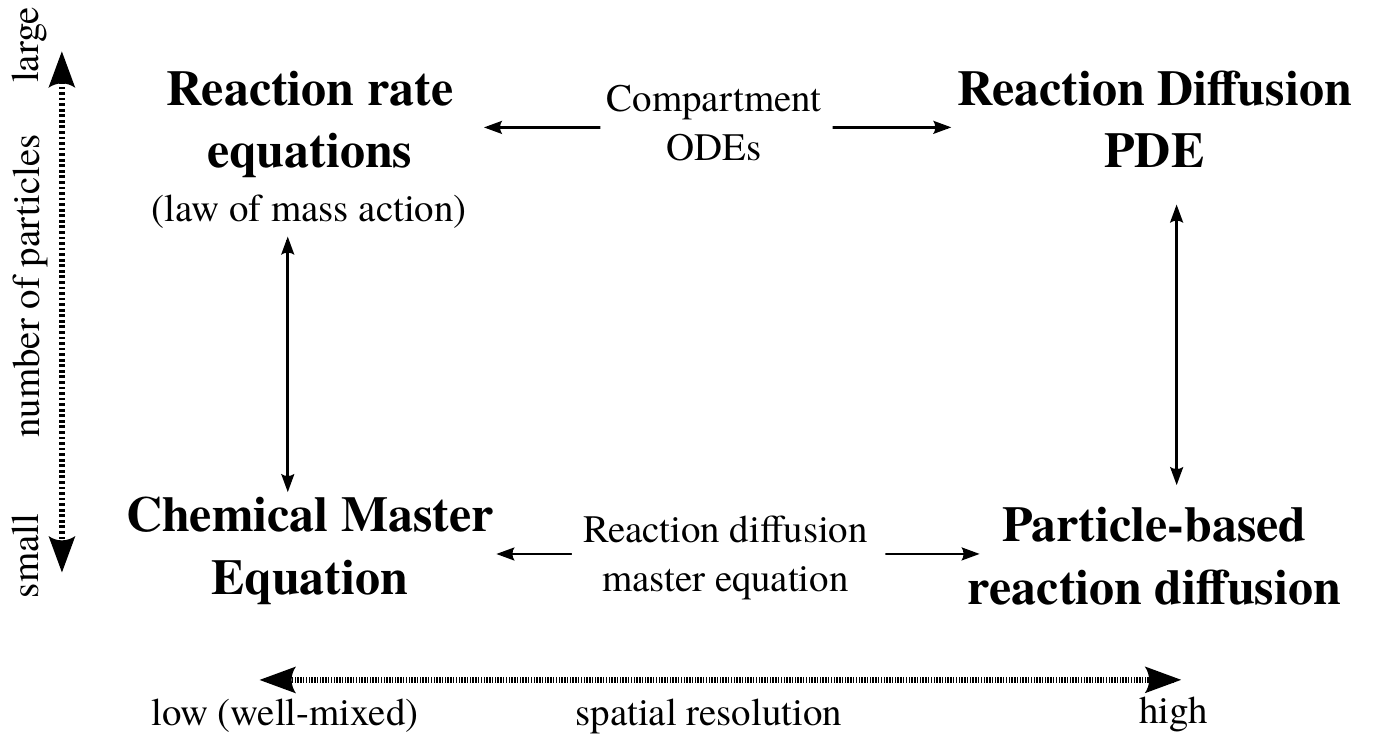}
		\caption{Some models of reaction-diffusion processes organized by their spatial
			scaling and the number of particles. Only the most relevant models for this work are shown here.}
		\label{fig:diagRDmodels}
	\end{figure}

	Reaction-diffusion processes take different forms depending on the number of particles involved and the scale of interest (Fig. \ref{fig:diagRDmodels}). When the number of particles is small, the stochastic fluctuations due to diffusion and chancy reactions need to be taken into account with a probabilistic particle-based approach. However, if the number of particles is large, small fluctuations in the number of particles become negligible, and deterministic concentration dynamics in the form of reaction-diffusion PDEs become a more suitable alternative. 
	
	In the case of well-mixed systems, the spatial component of these models is averaged out yielding either the well-known rate equations or the chemical master equation, depending on the number of particles (Fig. \ref{fig:diagRDmodels}). In this work, we concentrate on spatially inhomogeneous systems at different scales. We develop theoretical and simulations techniques that bridge PBRD simulations with reaction-diffusion PDEs, in particular in the context of open systems. We give below a brief overview of the two relevant reaction-diffusion models.
	
	\subsection{Particle-based reaction-diffusion}\label{sec:PBA}
\begin{figure}[ht]
		\center
		\includegraphics[width=\columnwidth]{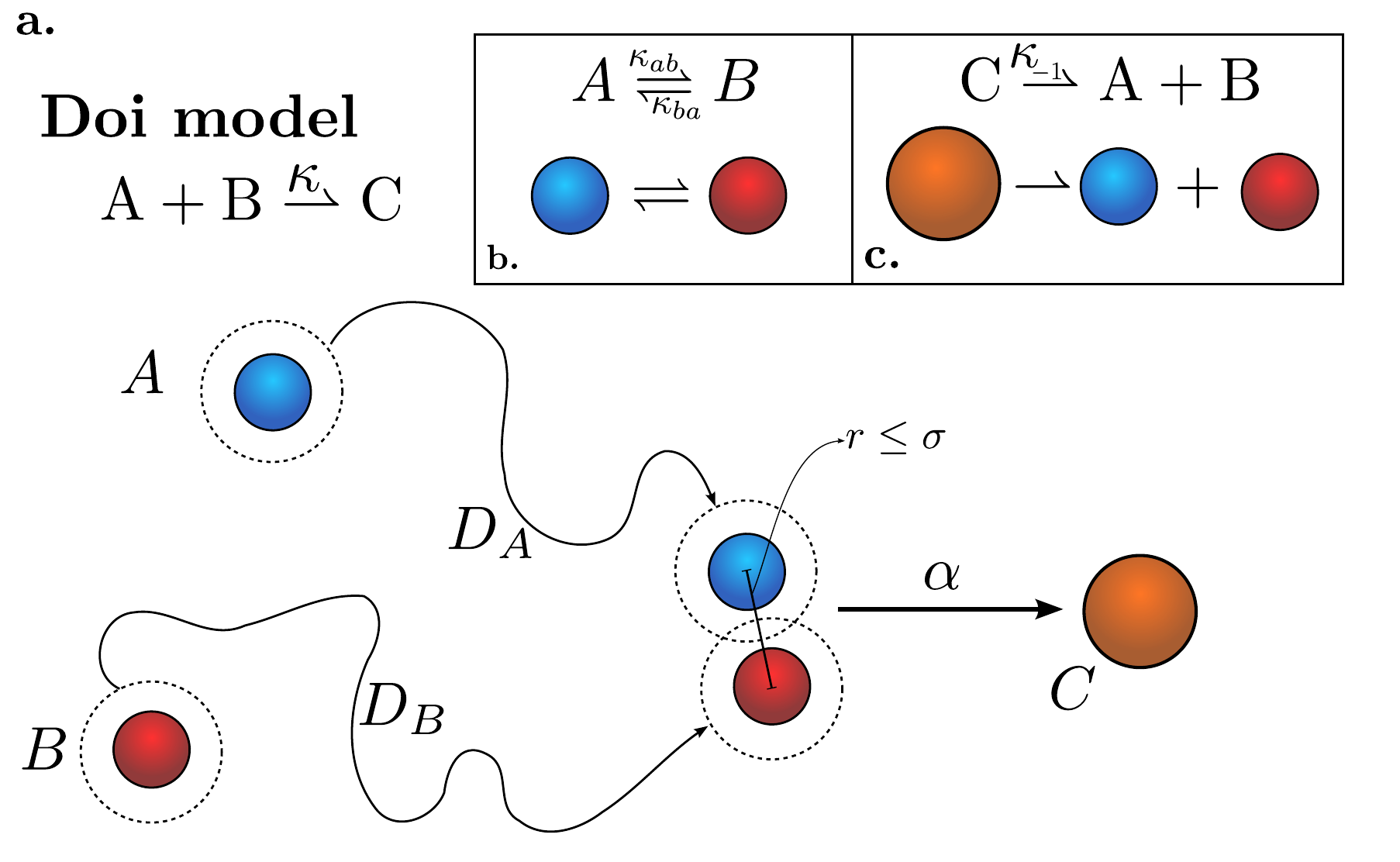}
		\caption{Illustration of some of the possible reactions in a particle-based reaction-diffusion simulation. \textbf{a.} Diagram of the Doi model for bimolecular reactions. When the two particles are closer than a distance of $\sigma$, they react with rate $\alpha$. Note it is not the same as the macroscopic rate $\kappa$. \textbf{b.} An example of a unimolecular reaction (first-order) reaction, where $A$ simply transforms into $B$. \textbf{c.} The backward reaction of the binding given by the Doi model. The products should be placed uniformly at a distance $\delta r$ such that $\delta r \leq \sigma$ to satisfy detailed balance \cite{frohner2018reversible}.}
		\label{fig:pbrd}
	\end{figure}

	The particle-based approach to model reaction-diffusion follows the approach used in the ReaDDy2 software \cite{hoffmann2019readdy} (other well-known software packages are \cite{andrews2010detailed,moraru2008virtual,stiles2001monte}). It consists on simulating each molecule as an spherical particle. The Brownian diffusion of each molecule is modeled using overdamped Langevin dynamics,
	\begin{align}
	dx(t) = \sqrt{2D}dw(t),
	\label{eq:odampedLang}
	\end{align} 
	where $x(t)$ is the position of the molecule at time $t$ and $w(t)$ is a collection of independent
	Wiener processes (one per coordinate). Reactions are modeled depending on the type of reaction (Fig. \ref{fig:pbrd}):  
	\begin{itemize}
		\item Zeroth order reactions, $\emptyset \xrightharpoonup[]{\kappa_0} A$: a new $A$ molecule is placed uniformly in the whole domain with rate $\kappa_0$.
		\item First order reactions, $ A \xrightharpoonup[]{\kappa_{\text{ab}}} B$: molecule $A$ is transformed into $B$ with rate $\kappa_\text{ab}$.
		\item Second order reactions, $ A + B \xrightharpoonup[]{\kappa} C$: if the relative distance between $A$ and $B$ is less than the reaction radius $\sigma$, the particles react with rate $\alpha$. The new molecule $C$ is placed in an averaged position between $A$ and $B$. 
	\end{itemize}
	The mathematical model for zeroth and first-order reactions is the same as in the well-mixed case, where there is no spatial dependence. Higher-order reactions can be decomposed into several second-order reactions, so the theory for second-order reactions results the most relevant. The mathematical theory we use to model second-order reactions is based on diffusion-influenced reactions, specifically in the Doi volume reactivity model (Fig. \ref{fig:pbrd}a) \cite{doi1976stochastic, teramoto1967theory}.
	
	The Doi model consists of an isolated pair of molecules $A$ and $B$, where $A$ is fixed at the origin, $B$ is a distance $r$ from $A$, and it undergoes Brownian diffusion (Eq. (\ref{eq:odampedLang})) with diffusion coefficient $D$. Further, $B$ can only react with $A$ with rate $\alpha$ if within the reaction radius $r\leq \sigma$. The probability of finding $B$ at distance $r$ at time $t$, providing it started at a distance $r_0$, is $f(r,t|r_0)$, which obeys the following Fokker-Planck equation
	\begin{align*}
	\partial_{t}f(r,t|r_0) = D \nabla^2 f(r,t|r_0) - \chi_{r\leq \sigma}(r) f(r,t|r_0),
	\end{align*} 
	where $\chi_{r\leq \sigma}(r)$ is the indicator function (1 if $r<\sigma$ and 0 otherwise). Note as the steady state is reached ($t\rightarrow \infty$), $f(r,t|r_0)$ goes to zero since the particle $B$ will react with probability 1. Following generalizations of diffusion-influenced reactions to reversible reactions \cite{agmon1984diffusion, agmon1990theory, cavallo2019reversible, del2014fluorescence, del2016discrete, frohner2018reversible, gopich2002kinetics, khokhlova2012comparison, kim1999exact}, we can also model the reversible reaction $C\xrightharpoonup[]{} A + B$, by placing the reactants uniformly within a distance $\sigma$ of each other. This choice ensures detailed balance is satisfied. Alternative solutions exists when there is an interaction potential involved \cite{frohner2018reversible} or when numerical efficiency is a priority \cite{andrews2004stochastic}. 
	
	Note the parameters required for a particle-based reaction-diffusion simulation are the microscopic rates and the reaction radius for second-order reactions.

	\subsection{Reaction diffusion PDEs}
	When a systems has a very large number of particles, its chemical kinetics are better described by deterministic concentration-based approach. Consider $c(t)$ the vector of concentrations of $N$ chemical species, which are involved in $M$ different reactions. The kinetics are given in terms of the following PDE,
	\begin{align*}
	\partial_{t} c= D\nabla^2 c + R(c),
	\end{align*}
	where here $D$ is a diagonal matrix with the diffusion coefficient of each species and $R(c)$ encapsulates all the $M$ reactions. This equation without the diffusion term would be of the form of the well-known law of mass action \cite{beard2008chemical}. As an example, consider the predator-prey dynamics
	\begin{align}
	A  \xrightharpoonup{\kappa_1} 2 A, \ \ \ \ \ \ \ \ \ 
	A + B  \xrightharpoonup{\kappa_2} B, \ \ \ \ \ \ \ \ \
	B  \xrightharpoonup{\kappa_3} \emptyset,
	\end{align}
	where $A$ represent the preys and $B$ the predators. The concentration dynamics are given by the Lotka-Volterra equations
	\begin{align*}
	\partial_t c_A = D_A\nabla^2 c_A + \kappa_1 c_A - \kappa_2 c_A c_B, \\
	\partial_t c_B = D_B\nabla^2 c_B + \kappa_2 c_A c_B - \kappa_3 c_B,
	\end{align*}
	where $c_A$ and $c_B$ represent the concentration of predators and preys, respectively. Naturally, initial and boundary conditions need to be provided to close the system.
	
	Reaction-diffusion PDEs, like the one just presented, can be solved using standard numerical methods, like finite difference schemes, finite elements and spectral methods. In this work, we use finite differences \cite{leveque2007finite, kostrethesis,salsa2013primeR} since they are simple and fit the purpose of this work. We mainly use the Crank-Nicolson method combined with an operator splitting approach if necessary, see Appendix \ref{sec:AppendixA} for brief implementation details.

	\section{Coupling particle-based models to chemical reservoirs}\label{sec:coupling}
\begin{figure}
		\textbf{a.} \includegraphics[width=0.9\columnwidth]{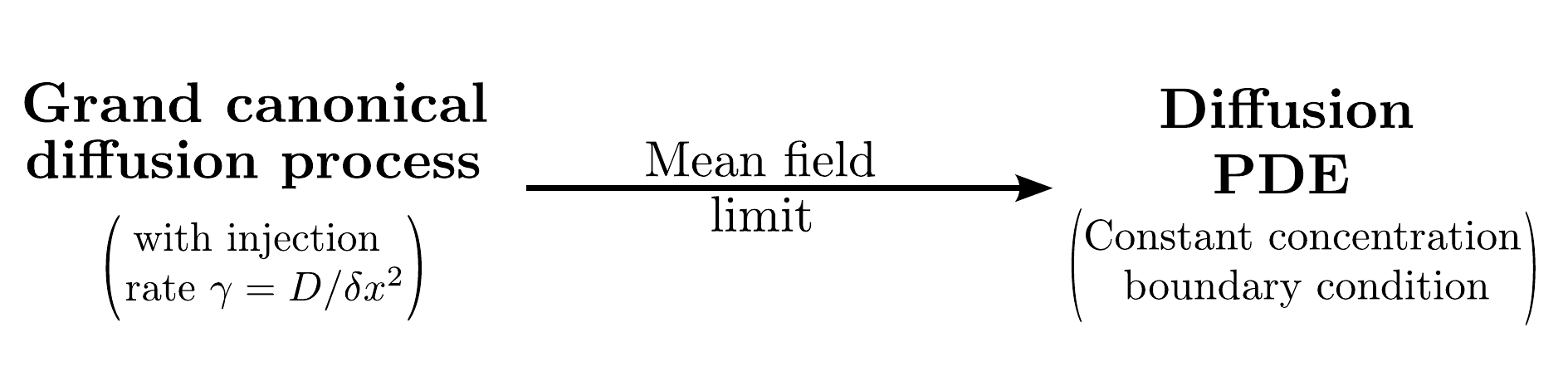} \\
		\textbf{b.} \includegraphics[width=0.9\columnwidth]{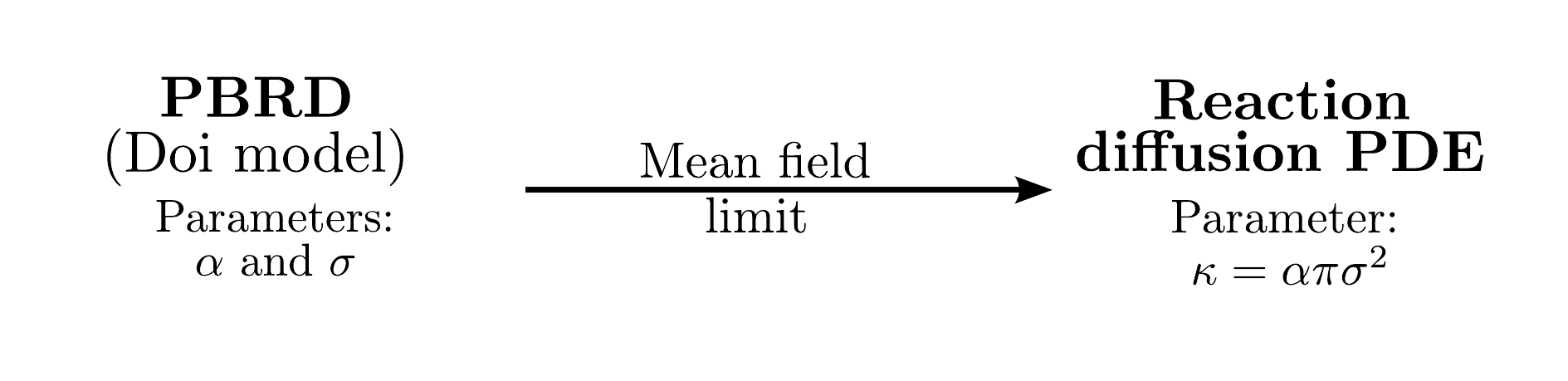}
		\caption{Diagram summarizing the two main results of Section \ref{sec:coupling}. \textbf{a.} Mean field limit of the particle-based diffusion open system. If the injection rate of particles from the reservoir is set to $\gamma = D/\delta x^2$ in the master equation, the mean field yields a constant concentration boundary condition. \textbf{b.} Mean field limit of PBRD. It relates the miscroscopic parameters of the Doi model, $\alpha$ and $\sigma$, with the macroscopic reaction rate $\kappa$. }
		\label{fig:meanFields}
	\end{figure}

	In this section, we derive two results that are fundamental to couple PBRD simulations with reservoirs mediated by reaction-diffusion PDEs. These results address the two main challenges mentioned in the introduction, and they are summarized in Fig. \ref{fig:meanFields}. 
	
	The first result determines how to consistently couple a particle-based model to a constant concentration reservoir. It does so by matching the mean-field limit of a particle-based diffusion process to its corresponding macroscopic PDE description. In this setting, the particle-based model is in contact with a chemical reservoir, so we call it a grand canonical diffusion process. In a simulation context, this result can be easily extended to reservoirs with spatially and time-dependent concentrations given by a reaction-diffusion PDE. 
	
	The second result shows how reaction-diffusion PDEs can be recovered as the mean-field of particle-based reaction-diffusion processes. This is essential to develop consistent coupling numerical schemes since it determines the relation between the microscopic and macroscopic parameters.
	
	
	\subsection{Mean-field of grand canonical diffusion processes} \label{sec:opendiffMF} 
	We want to consistently couple PBRD simulations to PDE mediated chemical reservoirs. To parametrize the coupling, we need to match the mean-field dynamics of particle-based models with its corresponding macroscopic PDE behavior.
	
	We begin with a one-dimensional system with an arbitrary number of noninteracting diffusing particles and a coupling to a constant concentration reservoir on one end; this is a grand canonical diffusion process. In the macroscopic setting, this corresponds to the diffusion PDE
	\begin{align}
	\partial_{t} c(x,t)= D\nabla^2 c(x,t), \label{eq:diff0}\\ 
	\partial_x c(x, t) |_{x=0} = 0  \quad c(R,t)=c_R,
	\nonumber
	\end{align}
	with $c_R$ a constant. The boundary condition at $x=0$ is not really relevant for our purpose, but we assume Neumann for simplicity.
	
\begin{figure}
		\includegraphics[width=0.9\columnwidth]{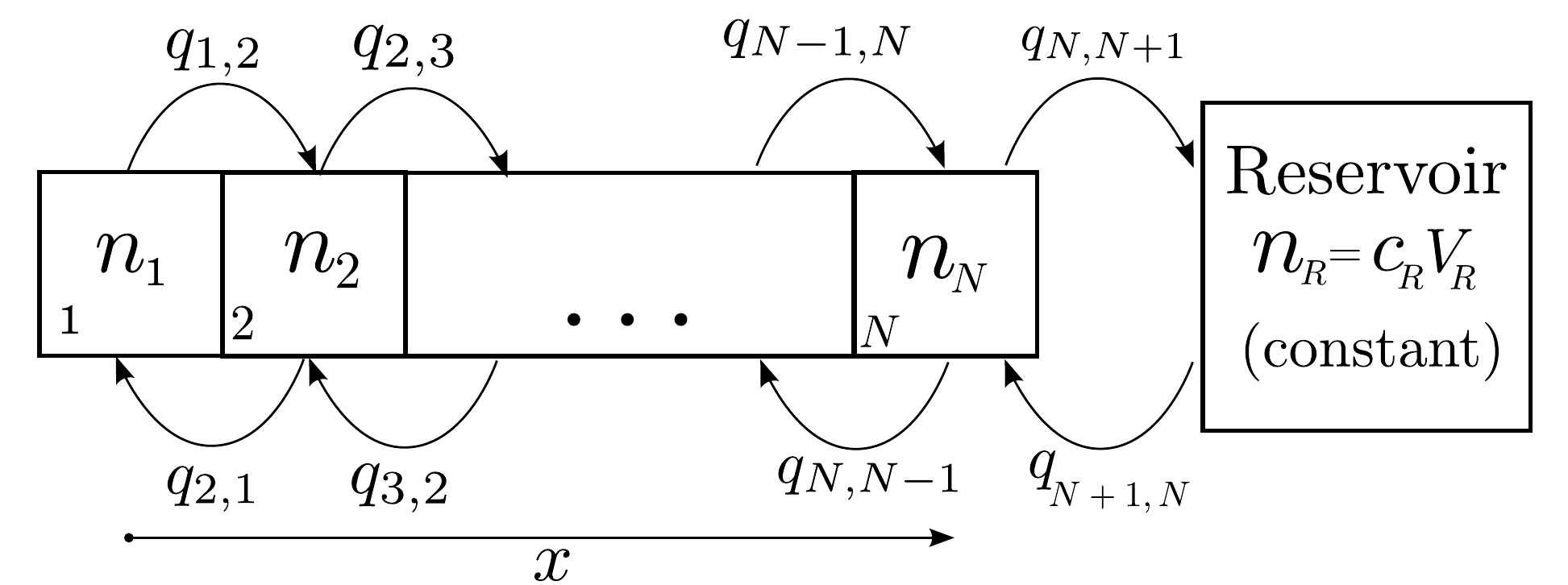}
		\caption{Diagram of the grand canonical master equation for an open system. It allows for an arbitrary number of diffusing particles, and it is coupled to a material reservoir. This is the result of discretizing the one-dimensional particle-based diffusion processes in contact with a constant concentration reservoir.}
		\label{fig:diagDiscret}
	\end{figure}

	In the particle-based setting, this corresponds to particles diffusing independently following standard Brownian motion. The reservoir on one end can absorb and introduce new particles into the system with a certain rate. To obtain the mean-field of the particle-based system, it will be convenient to discretize the domain $[0,R]$ into $N$ cells of size $\delta x$. The number of particles in the $i^{\text{th}}$ cell is denoted by $n_i$. The state of the systems is given by $P(n_1,\dots,n_N,t)$, which corresponds to the probability of having $\{n_1,\dots,n_N\}$ particles in the cells $\{1\dots,N\}$ at time $t$. We further assume cell $N$ is in contact with a reservoir in cell $N+1$ of volume $V_R$ with a constant concentration of particles $c_R$ at all times (Fig. \ref{fig:diagDiscret}b.). As the particles diffuse independently, the jump rates of each particle from cell $i$ to neighboring cell $j$ are simply the diffusion jump rates \cite{del2018grand},
	\begin{align}
	q_{i,j}=\frac{D}{\delta x ^{2}}, \ \ \ i\neq j.
	\label{eq:SMErates}
	\end{align}
	We can now write a master equation for the dynamics of $P(n_1,\dots,n_N,t)$ \cite{del2018grand},
	\small
	\begin{gather}
	\frac{d P(n_1,\dots,n_N,t)}{dt} = -P(n_1,\dots,n_N,t)\sum_{i=1}^N[q_{i,i+1} + q_{i,i-1}]n_i \nonumber\\
	+ \sum_{i=1}^{N-1} \bigg[
	P(n_i + 1, n_{i+1} -1) q_{i,i+1} (n_i + 1) + \label{eq:gcme} \\
	P(n_i - 1, n_{i+1} + 1) q_{i+1,i} (n_{i+1} + 1) \bigg] + \nonumber\\
	P(\dots, n_N + 1) q_{N,N+1} (n_N + 1)
	+ P(\dots, n_N - 1) q_{N+1,N} n_R, \nonumber
	\end{gather}
	\normalsize
	where we used $P(n_i + 1, n_{i+1} -1) = P(\dots, n_i + 1, n_{i+1} -1 \dots)$ to simplify notation. We refer to this equation as the grand canonical master equation since it describes an open system that can exchange particles with its environment. The terms in the first sum of Eq. (\ref{eq:gcme}) correspond to transitions that leave the current state; the second sum corresponds to transitions into the current state; and the last two terms corresponds to the transitions into the current state due to interactions with the reservoir. In order to recover the Neumann boundary condition at $x=0$, we set $q_{1,0}=0$, see Fig. \ref{fig:diagDiscret} and Eq. (\ref{eq:gcme}) for reference. Note the jump rate of particles from the reservoir into the system $\gamma=q_{N+1,N}$ is not yet known. We will refer to $\gamma$ as the injection rate.
	
	We will next study the continuous limit of the mean-field of Eq. (\ref{eq:gcme}) and obtain the value of the injection rate $\gamma$. The mean-field of the Eq. (\ref{eq:gcme}) is given by
	\begin{align}
	\sum\limits_{\{\bar{n}\}}n_{i}\frac{dP(n_1,...,n_N,t)}{dt}:=\frac{d \mean{n_{i}}}{dt},
	\end{align}
	where $\mean{n_{i}}$ is the expected number of particles at cell $i$. After some algebra, we obtain the following equation \cite{del2018grand,heuett2006grand} 
	\begin{align}
	\frac{d \mean{n_{i}}}{dt}= \mean{n_{i+1}} q_{i+1,i}- \mean{n_{i}} [q_{i,i+1}+q_{i,i-1}]+ \mean{n_{i-1}} q_{i-1,i}.
	\label{eq:meanGcme}
	\end{align}
	Writing everything in terms of concentrations $c_i=\mean{n_{i}}/\delta x$ and substituting the rates, we can take the limit $\delta x \rightarrow 0$. Note that the concentration remains bounded because $\mean{n_{i}}$ goes to zero at the same rate as the volume shrinks. This yields \cite{del2018grand}
	\begin{align}
	\partial_{t} c(x,t)=D\nabla^2 c(x,t),
	\label{eq:diff}
	\end{align}    
	which is not surprisingly the diffusion equation for the concentration $c(x,t)$. It is also straightforward to check, we recover the Neumann boundary condition at $x=0$ by using a ghost cell approach \cite{leveque2007finite}. However we need to be careful at the boundary in contact with the reservoir. We denote $F_i:=\mean{n_i}/\delta x$ the mean concentration at cell $i$. We write Eq. \ref{eq:meanGcme} for $i=N$ and substitute the corresponding rates
	\begin{align*}
	\frac{dF_{N}}{dt}=q_{N+1,N}c_R+\frac{D}{\delta x^{2}}(-2F_{N}+F_{N-1}).
	\end{align*}
	We add a ghost cell $N+1$ that represents the reservoir. As the concentration in the material bath is constant, we set $c_{R}=F_{N+1}$ and
	rewrite this equation as
	\begin{align*}
	\frac{dF_{i}}{dt}=\frac{D}{\delta x^{2}}(F_{N+1}-2F_{N}+F_{N-1})+ \\
	q_{N+1,N}c_R-\frac{D}{\delta x^{2}}F_{N+1}.
	\end{align*}
	The first term corresponds to the discretized diffusion equation (Eq. \ref{eq:diff0}) that we want to recover in the continuous limit. Therefore, the additional terms must be zero. As $c_R = F_{N+1}$, this implies that the injection rate is
	\begin{align}
	\gamma =q_{N+1,N} = \frac{D}{\delta x^2}.
	\label{eq:gammarate}
	\end{align}
	This is the jump rate of particles from the reservoir into the system, such that the macroscopic Eq. (\ref{eq:diff0}) with its boundary conditions are recovered in the continuous mean field limit. Not surprisingly, it matches the diffusion jump rate.
	
	This result establishes the connection between the particle-based and the concentration-based approach for open systems. It can be used to implement particle-based simulations in contact with material reservoirs with constant concentrations, see Section \ref{sec:hybridscheme}. It is also straightforward to extend it to more complicated systems, such as time and space-dependent reservoirs.
	
	Note we assumed that the concentration $c_R$ remains constant, even when extracting particles from the system. This is only possible if we make the number of particles in the reservoir and its volume both infinite while keeping the concentration constant. We implicitly make this assumption when taking the continuous limit.
	
	To implement a simulation of this process, we can discretize time at first order, so the jump rates of each particle (Eq. (\ref{eq:SMErates})) become jump probabilities $p_{i,j}=\Delta t D /\delta x ^{2}$, $i\neq j$ \cite{del2016discrete}. We can then use these, along with the jump probabilities of reservoir particles, to implement the simulation. However, as the probability of one particle jumping either left or right is at most one, then $2\Delta t D /\delta x ^{2}\leq 1$, which constrains the simulation time step $\Delta t$ to, 
	\begin{align}
	\Delta t \leq \delta x^2/(2D).
	\label{eq:maxdeltat}
	\end{align}
	This will be important for the implemenation of the Hybrid scheme in Section \ref{sec:hybridscheme}.
	
	The results of this section are based on \cite{del2018grand}, and they can also be obtained using the law of large numbers. We advise the reader to consult \cite{del2018grand} for additional details.

	\subsection{Mean-field of particle-based reaction-diffusion processes} \label{sec:RDprocMF}
	In this section, we obtain the mean-field dynamics of particle-based reaction-diffusion processes. This result provides a mathematical bridge between the microscopic particle-based approach and the macroscopic deterministic approach for reaction-diffusion processes. It establishes the relation between the microscopic and macroscopic parameters. The diffusion coefficients and the reaction rates for up to first order reactions remain the same for particle-based models and macroscopic models \cite{del2018grand}. However, this is not the case for second-order reactions (bimolecular reactions). We thus focus on deriving this result for bimolecular reactions, 
	\begin{align}
	A+B \rightarrow C.
	\end{align}
	We denote $\alpha$ as the microscopic reaction rate based on the Doi model (Section \ref{sec:PBA}). We want to determine the relation between the particle-based simulations of this reaction with the corresponding macroscopic deterministic reaction-diffusion PDE
	\begin{align}
	\partial_t c_A = D_A \nabla^2 c_A - \kappa c_A c_B, \label{eq:exRDPDE} \\
	\partial_t c_B = D_B \nabla^2 c_B - \kappa c_A c_B, \nonumber
	\end{align}
	where $c_A$ and $c_B$ denote the deterministic concentrations of $A$ and $B$, and $\kappa$ is the macroscopic reaction rate. This will naturally provide a connection between $\alpha$ and $\kappa$. We obtain this result for two dimensions, but it extends naturally to higher dimensions.
	
	We begin with the particle-based description. Once again, it is convenient to discretize the two dimensional domain, $\Omega$, in $n$ square cells denoted by $V_{1},V_{2},....,V_{n}$ of length and height $h$. Denote by $X^{A}_{i}(t)$ and $X_{i}^{B}(t)$ the number of $A$ and $B$ particles in cell $i$ at time $t$. Following the Doi model for particle-based reactions \cite{doi1976stochastic, hoffmann2019readdy}, we assume $A,B$ react with rate $\alpha$ if they are are closer than a distance $\sigma$. Using the reaction diffusion master equation formalism \cite{isaacson2013convergent}, the change in number of $A$ particles in cell $i$ can be written using the Kurtz representation \cite{anderson2015stochastic}
	\begin{align}
	X^A_i(t) =& X^A_i(0) - \sum^n_{j=1} 
	U_j\left( \int_0^t \alpha \phi_{ij} X^A_i(s) X_j^B(s) ds \right) \nonumber \\
	&+ \mathcal{D}_i^t(X^A),
	\label{eq:numAtraj}
	\end{align}
	where $X^A=\{X_1^A,\dots X_N^A\}$ and the term $\mathcal{D}_i^t$ denotes the discrete change of number of particles at cell $i$ due to diffusion after a time $t$. This operator can be explicitly written in terms of the discrete Laplace operator and the elapsed time $t$. Each $U_j$ denotes a unit rate Poisson process, where the rate function $\int_0^t \lambda (s)ds$ depends on the propensity $\lambda (s) = \alpha \phi_{ij} X^A_i(s) X_j^B(s)$ of the reaction. We can analogously formulate an equation for $X_{i}^{B}$ and do the same process as below. 
	
	Note the reaction propensity not only depends on the microscopic rate $\alpha$ but also on the quantity $\phi_{ij}$. This will help us quantify how likely the reaction is to happen, given that the $A$ particles are in cell $V_i$ and $B$ particles are in cell $V_j$. This is defined as
	\begin{align} 
	\phi_{ij}= \frac{\vert \mathcal{R}\cap V_{ij} \vert}{\vert V_{ij} \vert},
	\label{eq:defPhiij}
	\end{align}
	where $\mathcal{R}$ is the reactive region in the 4-dimensional space defined by positions $x$ and $y$ such that $|x-y|\leq \sigma$, $V_{ij}$ is the hypercube formed by $V_i \times V_j$ with volume $h^4$ and the bars denote we are taking the volume of these regions. This quantity is simply the ratio between the reactive volume contained within $V_{ij}$ and the total volume of $V_{ij}$. If $A$ particles are well-mixed in $V_{i}$ and $B$ particles are well-mixed in $V_{j}$, it gives the probability of $A$ and $B$ particles being close enough to react, see \cite{isaacson2013convergent} for more details on this quantity.
	
	Rewriting Eq. (\ref{eq:numAtraj}) in terms of concentrations, we obtain
	\begin{align}
	C^A_i(t) =& C^A_i(0)-\frac{1}{ \vert V_i \vert}   \sum_{j=1}^n  U_j\left( \int_0^t\alpha\phi_{ij} \vert V_{ij} \vert C^A_i(s) C^B_j(s) ds \right)
	\nonumber \\
	&+ \mathcal{D}_i^t(C^A),
	\label{eq:concAtraj}
	\end{align}
	where we used the linearity of the diffusion operator.
	Note the rate of the unit Poisson process depends on the concentrations, which are random variables themselves. These are therefore \textit{doubly-stochastic} processes, and they are called mixed Poisson processes or more generally Cox processes \cite{grandell1997mixed, schnoerr2016cox}. The expected value of a mixed Poisson process corresponds to the expected value of the random rate function \cite{grandell1997mixed}. Therefore the expected value of each of the Poisson processes $U_j$ from Eq. (\ref{eq:concAtraj}) is
	\begin{align*}
	\int_0^t\alpha\phi_{ij} \vert V_{ij} \vert  \langle C^A_i(s) C^B_j(s)\rangle ds.
	\end{align*}
	In general, $\langle C^A_i C^B_j\rangle = \langle C^A_i\rangle \langle C^B_j\rangle + \mathrm{Cov}(C^A_i, C^B_j)$ since they are not independent. Assuming the number of particles is large enough, as the particles are only correlated through reactions limited to a small reaction volume, the covariance is negligible in comparison to the product of the means. We can use these results to calculate the expected value of equation (\ref{eq:concAtraj}) 
	\begin{align}
	\langle C^{A}_{i}(t) \rangle =& \langle C^{A}_i(0)\rangle-\sum_{j=1}^n
	\vert V_{j} \vert\int_{0}^{t}\alpha \phi_{ij} \langle C^A_i(s)\rangle \langle C^B_j(s)\rangle ds
	\nonumber
	\\
	&+ \mathcal{D}_i^t(\langle C^A \rangle),
	\label{eq:concAtraj2}
	\end{align}
	where we used the linearity of the discrete diffusion operator to pass the expectation into the argument. We further derive with respect to $t$ both sides of the equation, yielding 
	\begin{align}
	\frac{d\langle C^{A}_{i}\rangle}{dt}=\mathcal{D}_i(\langle C^A \rangle)
	-\sum_{j=1}^n \alpha \phi_{ij}\vert V_j \vert \langle C^A_i\rangle \langle C^B_j\rangle,
	\label{eq:AexpectedConc}
	\end{align}
	
	where $\mathcal{D}_i$ is now the time-continuous diffusion operator in terms of rates instead of jump probabilities. 
	
	Equation (\ref{eq:AexpectedConc}) is still discrete in space; we now take the continuous limit $h\rightarrow 0$. In this limit, the first term converges to the continuous diffusion operator $\mathcal{D}$, see \cite{del2018grand, wang2003robust} for details. The limiting behavior of the second term is not trivial, since we need to solve
	\begin{align}
	\lim_{h\rightarrow 0} \left( \sum_{j=1}^n 
	\alpha \phi_{ij}\vert V_j \vert \langle C^A_i\rangle \langle C^B_j \rangle \right),
	\label{eq:limitRate}
	\end{align}    
	where $\phi_{ij}$, as defined in Eq. (\ref{eq:defPhiij}), can be expressed as the following integral
	\begin{align}
	\phi_{ij}=\frac{1}{\vert V_{ij}\vert}\int_{V_{i}}\int_{V_{j}}\chi_{\vert x-y\vert <\sigma} dydx,
	\label{eq:phieq}
	\end{align}
	where $\chi$ is the indicator function.
	If our domain is  a two-dimensional cube, then $x$ and $y$, each correspond to 2-dimensional vectors, and we have an integral over a hypercube. Substituting Eq. (\ref{eq:phieq}) into Eq. (\ref{eq:limitRate}), we obtain
	\begin{align*}
	&\lim_{h\rightarrow 0} \left( \frac{\alpha}{h^2} \langle C^A_i\rangle \sum_{j=1}^n 
	\langle C^B_j \rangle \int_{V_i}\int_{V_j}\chi_{\vert x-y\vert <\sigma} dydx \right),
	\end{align*}
	where we used that $\vert V_i \vert = h^2 =$, $\vert V_{ij} \vert = h^4$. Rearranging the terms, we obtain
	\begin{align*}
	=&\lim_{h\rightarrow 0} \left( \frac{\alpha }{h^2} \langle C^A_i\rangle \int_{V_i}
	\sum_{j=1}^n \langle C^B_j \rangle\int_{V_j}\chi_{\vert x-y\vert <\sigma} dydx \right),    
	\end{align*}
	As $\sigma$ is small, we can assume the value $\langle C^B_j \rangle$ does not change much in the domain of interest, $\vert x-y\vert <\sigma$, around cell $i$. We thus approximate it by its central value $\langle C^B_i \rangle$. Using this approximation, we can combine the sum over $j$ and the integral over $V_j$ into one integral over the whole domain $\Omega$, yielding
	\begin{align*}
	= & \lim_{h\rightarrow 0} \left(\frac{\alpha}{h^2}  \langle C^A_i\rangle \langle C^B_i \rangle\int_{V_i} \int_{\Omega}\chi_{\vert x-y\vert <\sigma} dydx \right).
	\end{align*}
	We further approximate the integral of the indicator function over $\Omega$ by an integral on the whole 2-dimensional space. This approximation is exact everywhere except in a small region close to the boundaries of $\Omega$. Thus, it yields the area of a circle of radius $\sigma$. 
	\begin{align*}
	= & \lim_{h\rightarrow 0} \left(\frac{\alpha }{h^2} \langle C^A_i\rangle \langle C^B_i \rangle  \pi \sigma^2 \int_{V_i}  dx \right),
	\end{align*}
	The remaining integral is simply $h^2$, so
	\begin{align*}
	=\lim_{h\rightarrow 0} \left(\alpha \pi \sigma^2 \langle C^A_i\rangle \langle C^B_i \rangle  \right) 
	= \alpha \pi \sigma^2 c_A(x_i) c_B(x_i),
	\end{align*}
	where $c_A = \langle C^A \rangle$ and $c_B = \langle C^B \rangle$ are continuous functions in space and time.
	We need to be careful with this limit since the indexing can change as $h\rightarrow$. Here we assumed the mean concentrations $\langle C^A_i\rangle$ and $\langle C^B_i \rangle$ are always centered in a fixed value $x_i$ as $h\rightarrow 0$. Note this result is only an approximation, albeit a very accurate one for $\sigma \ll \Omega$. With this result, the continuous space limit of Eq. (\ref{eq:AexpectedConc}) is then the familiar reaction diffusion PDE 
	\begin{align*}
	\frac{\partial c_A }{\partial t}= \mathcal{D}(c_A) - \alpha \pi \sigma^2  c_A c_B.
	\end{align*}
	Note the same equation can be obtained for the kinetics of $c_B$. The operator $\mathcal{D}$ corresponds to the well-known diffusion operator and comparing to Eq. (\ref{eq:exRDPDE}), the macroscopic rate $\kappa$ is simply
	\begin{align}
	\kappa = \alpha \pi \sigma^2.
	\label{eq:birateRel}
	\end{align}
	This relation holds for two dimensions, and it is of the form $\kappa = \alpha V_\text{react}$, where $V_\text{react}$ is the volume of the reactive volume. In three dimensions, we can analogously show this relation has the same form with the corresponding reactive volume
	\begin{align*}
	\kappa =  \alpha 4\pi \sigma^3/3.
	\end{align*}
	The result we just derived shows the reaction-diffusion PDE is the mean-field of the particle-based simulation based on the Doi model, assuming the number of particles is sufficiently large (otherwise covariances need to be taken into account). It also provides a connection between the microscopic bimolecular reaction rate $\alpha$ and the reaction radius $\sigma$ with the macroscopic reaction rate $\kappa$. As a side note, it is interesting that the same result is obtained when $\sigma \ll h$. Also note the diffusion coefficient does not play a role in this relation. 
	
	We completely neglected the covariances when taking the expectation of the mixed Poisson process from Eq. (\ref{eq:concAtraj}). Taking into account the covariances would yield a modified mean-field behavior valid at mesoscopic scales. The authors are currently working on a formal and more general version of this result that takes covariances into account \cite{kostreHydrolimit}.
	
	%
	
	\section{Hybrid scheme}\label{sec:hybridscheme}
	In this section, we use the results from Section \ref{sec:coupling} to derive a hybrid scheme to couple particle-based simulations with reservoirs mediated by reaction-diffusion PDEs. 
	
\begin{figure*}
		\flushleft
		\textbf{a.} 
		\includegraphics[width=0.95\columnwidth]{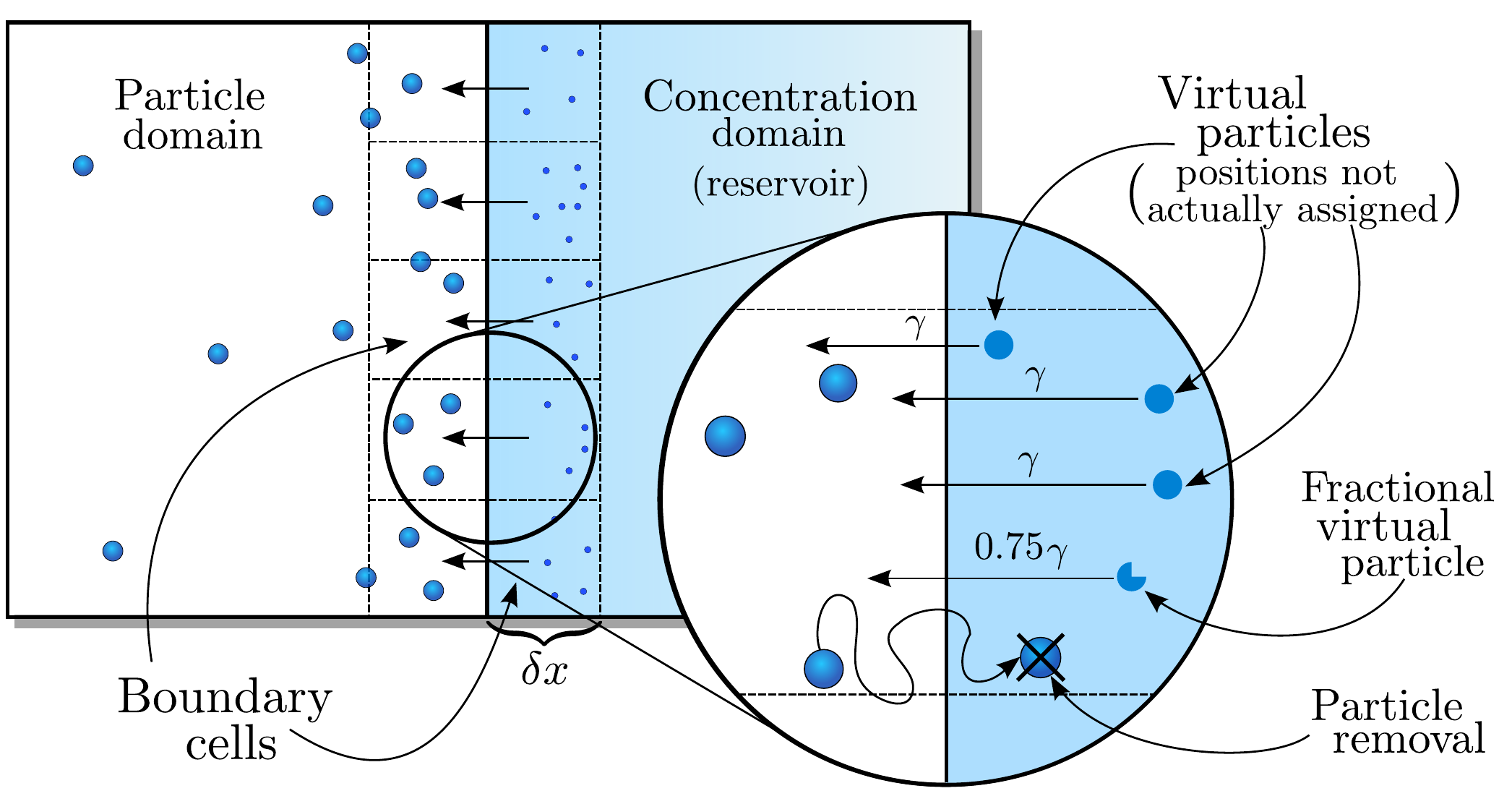} 
		\textbf{b.} 
		\includegraphics[width=0.95\columnwidth]{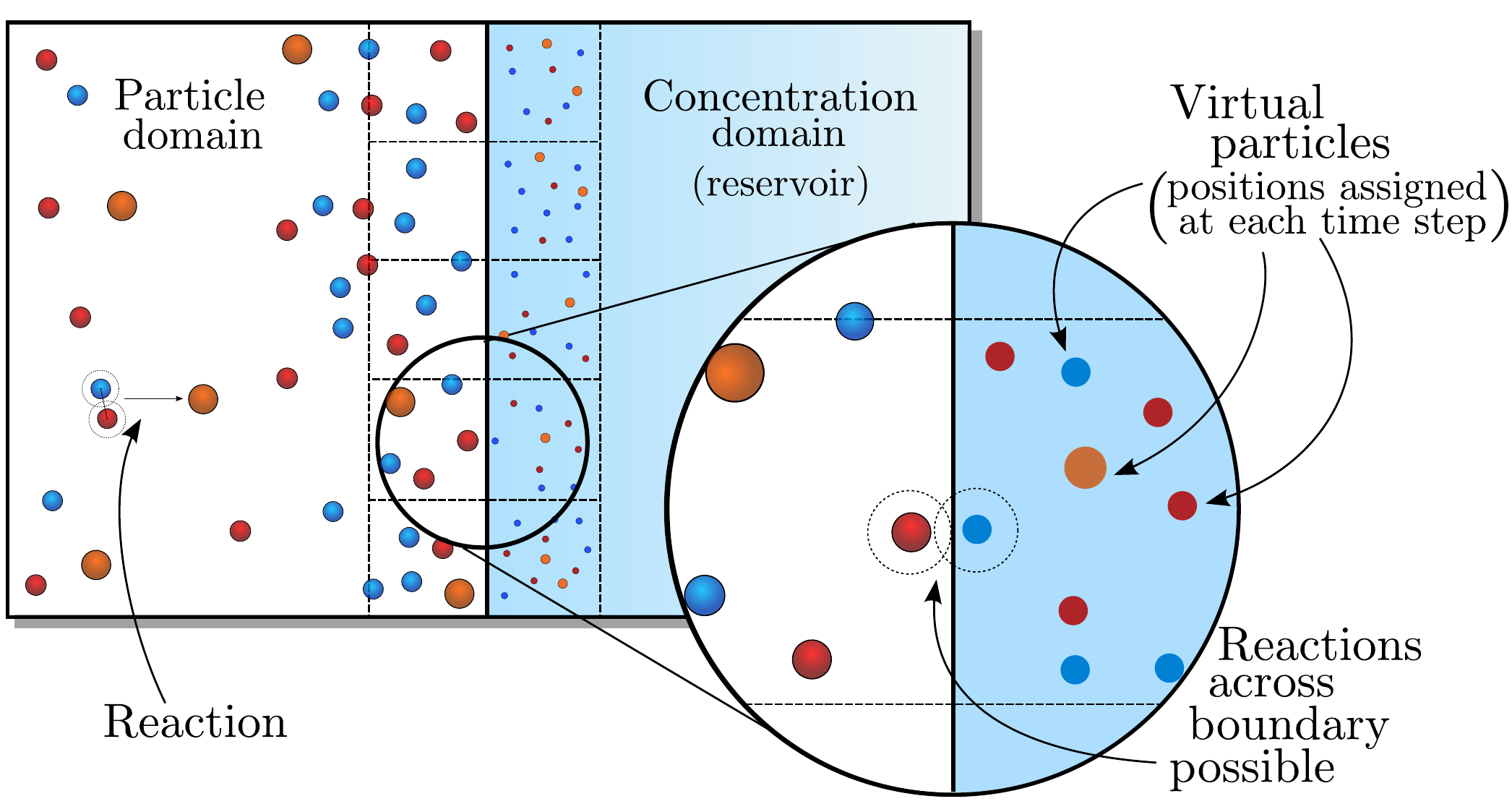}
		\caption{Illustrations of the boundary coupling in the hybrid scheme. The domain is divided by an interface into the particle and the concentration domain. \textbf{a.} Boundary coupling for one diffusing species. In the particle domain, particles diffuse freely following Brownian motion. If a particle diffuses into the concentration domain (the reservoir), it is eliminated. Along the boundary cells in the concentration domain, we convert the concentration into particle number, generally a non-integer value. The integer part is the number of virtual particles, each can jump with rate $\gamma$ into the boundary cells in the particle domain. The fractional part corresponds to a \textit{fractional} virtual particle, whose jump rate is scaled by this fraction. Note virtual particles are only drawn for illustration purposes, and they do not have a specific position within the boundary cell. The same procedure applies for multiple species with up to unimolecular reactions. \textbf{b.} Boundary coupling for a system with three species $(A,B,C)$ involving a bimolecular reaction $A+B\rightarrow C$. The coupling is analogous to the one in Fig. \ref{fig:HybridSetting}a and follows the particle-based dynamics described in Section \ref{sec:PBA}. If a red particle (A) is close enough to a blue one (B), they can react, and the product (C) is placed in the average position between the two particles. Unlike the coupling from Fig. \ref{fig:HybridSetting}a, the positions of the virtual particles are sampled uniformly within the boundary cell. This allows for bimolecular reactions to occur within all boundary cells across the coupling boundary, which makes the coupling accurate and robust. If the position of the reaction product is within the concentration domain, it is eliminated. This coupling can be applied to a general system with an arbitrary number of species with up to second-order reactions.}
		\label{fig:HybridSetting}
	\end{figure*}

	The essence of the algorithm is illustrated in Fig. \ref{fig:HybridSetting}. The domain is split into the particle domain and the concentration domain. The dynamics on the concentration domain, which functions as the reservoir, are modeled by a reaction-diffusion PDE, which is solved either analytically or with standard finite difference methods. As this is well documented in the literature \cite{leveque2007finite,salsa2013primeR} (see Appendix \ref{sec:AppendixA}), we concentrate on the particle domain where the dynamics are governed by three processes: injection, reaction and diffusion. We describe next these processes and show how they are combined into one algorithm.
	
	\subsection{Injection}
	We use the result from Section \ref{sec:opendiffMF}, mainly Eq. (\ref{eq:gammarate}), to obtain a consistent scheme to inject particles from the reservoir into the particle system. Along the edge between the particle and the concentration domain, we create squared boundary cells of edge length $\delta x$ (Fig. \ref{fig:HybridSetting}a). Every time iteration and for every boundary cell $i$ in the concentration domain, we calculate the average concentration $c_i$ within the boundary cell and convert it into number of particles $N_i$ (multiplying by cell volume $V_i$). The resulting values will be in general non-integers. The integer part corresponds to the number of virtual particles, and the fractional part to one \textit{fractional} virtual particle. We call them virtual particles because they do not belong to the particle domain. Given this information, the injection of particles for a time-step $\Delta t$ follows the following procedure.
	
	\textbf{Injection procedure}
	When called for a time interval $\tau$, this procedure follows these steps:
	\begin{enumerate}
		\item Let each virtual particle jump into the corresponding neighboring boundary cell in the particle domain with the injection rate $\gamma$ (Eq. (\ref{eq:gammarate})). This corresponds to jumping with probability $1-\exp (-\gamma \tau)$.
		\item Let the fractional virtual particles jump in the same way with a rate $\gamma$ scaled by the corresponding fraction value. For example, if $N_i=15.87$, then the rate for the fractional virtual particle is $0.87\gamma$.
		\item For every successful jump event, place a new particle uniformly in the corresponding boundary cell of the particle domain.   
	\end{enumerate}
	
	\subsection{Reaction}
	Most reactions are or can be decomposed into unimolecular or bimolecular reactions, so we only focus on these types of reactions. We employ the methodology introduced in Section \ref{sec:PBA} to implement the particle-based simulation in the particle domain. We further use the result from Section \ref{sec:RDprocMF} to establish a relation between the parameters of the PBRD simulation and the parameters of the reaction-diffusion PDE. The diffusion coefficients and unimolecular reaction rates remain the same on both models, while the bimolecular reaction rate from the PDE can be obtained from PBRD parameters following Eq. (\ref{eq:birateRel}). 
	
	\textbf{Zeroth and first-order reactions}
	These reactions depend uniquely on zero or one particle, e.g. creation of particles and conformational changes, respectively. The diffusion coefficients of zeroth and first-order reaction rates are the same in the PBRD simulation as in the PDE. Given a reaction rate $k$, on the particle-based simulation, the probability of a reaction to happen within a time interval $\Delta t$ is given by
	\begin{align}
	1-\exp (-k\Delta t).
	\label{eq:probReact}
	\end{align}
	The boundary coupling remains the same as in the injection process, as shown in Fig. \ref{fig:HybridSetting}a.
	
	\textbf{Second order reactions}
	These reactions depend on two particles, so they are also called bimolecular reactions. Given a microscopic bimolecular reaction rate $\alpha$, we calculate analogously the probability of a reaction within a time interval $\Delta t$ as $1-\exp (-\alpha\Delta t)$, but we only do so if the reactants are within a distance $\sigma$ of each other. The microscopic bimolecular reaction rate needs to be consistent with the PDE reaction rate $\kappa$. Following Eq. (\ref{eq:birateRel}), the relation should be $\kappa = \alpha \pi \sigma^2$. 
	
	In order to incorporate bimolecular reactions, we need to modify the boundary coupling (Fig. \ref{fig:HybridSetting}b). Considering a reactant in the particle domain can react with another reactant in the concentration domain, we risk losing accuracy in the boundary region if we do not allow for these reactions to happen. We can solve this by applying the following steps at a given time step (Fig. \ref{fig:HybridSetting}b):
	\begin{enumerate}
		\item Uniformly sample the positions of the virtual particles within the boundary cells in the concentration domain. The fractional virtual particle is placed with a probability equal to the fraction value.
		\item Allow the particles in the particle domain, along with the particles and virtual particles in all the boundary cells to react.
		\item If a reaction happens, sample the location of the product, usually given by the average position of the reactants. If the product location is in the particle domain, place the new particle, otherwise eliminate it from the simulation. 
	\end{enumerate}
	Note the number and position of virtual particles are resampled at the beginning of each time step.
	
	\textbf{Order of reactions}
	There are several possible ways to deal with the order of reactions at a given time step. One possibility is to choose the reaction event by drawing the next reaction event uniformly from all possible events while avoiding conflicting events to happen simultaneously. In this work, we apply this approach in conjunction with a Strang splitting \cite{strangsplitting}, which leads to the reaction procedure. An alternative approach is to weight the probability of possible reaction events with their respective reaction probability, as done in ReaDDy 2 \cite{hoffmann2019readdy}.
	
	\textbf{Reaction procedure}
	When called for a time interval $\tau$, this procedure follows these steps: 
	\begin{enumerate}
		\item Sample all possible zeroth and first-order reactions happening within $\tau/2$. Select uniformly which reaction happens and apply them while avoiding conflicting events.
		\item Sample all possible second-order reactions happening within $\tau$. Select uniformly which reaction happens and apply them while avoiding conflicting events.
		\item Sample, select and apply again zeroth and first-order reactions happening within $\tau/2$.
	\end{enumerate}
	
	If there are no second-order reactions, there is no need to do the Strang splitting, and we can simply sample and apply all possible zeroth and first-order reactions for a time $\tau$. We've found this approach the most stable at the coupling boundary.  
	
	\subsection{Diffusion}
	All the particles in the particle domain, including the added ones during the injection and reaction steps, diffuse following standard Brownian motion. This can be simulated by applying the Euler-Maruyama scheme \cite{higham2001algorithmic} to the dynamics of each molecule (Eq. (\ref{eq:odampedLang})),
	\begin{align}
	x^{n+1}=x^n +\sqrt{2D\Delta t} \xi^n,
	\label{eq:eulermaruyama}
	\end{align}
	where $x^n$ is the position of one molecule at the $n^{th}$ time iteration. The time is $t=n\Delta t$ and $\xi^n$ is a vector with each entry sampled at every time step from a normal distribution with mean zero and variance one. The diffusion coefficient $D$ will be different for every chemical species. The dimensions of $\xi$ corresponds to the dimensionality of the problem (one, two or three). 
	
	If a particle diffuses from the particle domain into the concentration domain, it must be removed from the simulation. The desired boundary conditions along the other boundaries of the particle domain should be set.
	
	
	\subsection{Algorithm}
	\label{sec:algorithm}
	The process we are modeling is composed of several coupled processes: injection, reaction and diffusion, each of which can follow a different timescale. From a numerical standpoint, it is more stable, robust and accurate to integrate them using a Strang splitting \cite{del2018grand, strangsplitting}. In one time step $\tau$, the Strang splitting integration could be as follows: integrate injection and reactions for $\tau/2$, integrate diffusion for $\tau$ and integrate again injection and reactions for $\tau/2$. Unlike a straightforward integration, the Strang splitting allows for some newly created particles, due to injection or reactions, to diffuse and maybe even react again within the same time step, improving the modeling accuracy of the coupling between the processes within one time step. 
	
	In this algorithm, we implement two Strang splittings. The main Strang splitting integrates the injection, reaction and diffusion processes. The secondary one is implemented within the reaction procedure to smoothly integrate zeroth and first-order reactions with second-order reactions. The scheme is as follows: 
	
	\textit{Main input variables}: time iterations $N$, time step size $\Delta t$, boundary cell width $\delta x$, diffusion coefficients of all species involved and reaction parameters for all reactions considered.
	
	For every time step $t\in\{0, \Delta t,...,N\Delta t\}$:
	\begin{enumerate}
		\item For every species $X$ and every boundary cell $i$ in the concentration domain, calculate its average concentration $c_i^X$ and convert it into number of particles $N_i^X=c_i^X V_i$. Then, get the corresponding number of virtual and fractional virtual particles at each boundary cell. 
		
		\item For the species involving bimolecular reactions, sample uniformly the locations of the virtual particles within the boundary cells.
		
		\item Inject particles from the concentration domain into the particle domain for half a time step, $\Delta t /2$, following the injection procedure. Note that the injection rate depends on the diffusion coefficient, so it will be different for different chemical species.
		
		\item Determine and apply reactions in the particle domain occurring within half a time step, $\Delta t/2$, following the reaction procedure. This consists of a secondary Strang splitting, where zeroth and first-order reactions are applied for $\Delta t/4$, second-order reactions for $\Delta t/2$ and again first and zeroth order reactions for $\Delta t/4$. Note that for bimolecular reactions, the particles can react with the virtual particles in the boundary cells. If there are only zeroth and first-order reactions, the secondary Strang splitting is not necessary.
		
		\item Diffuse all particles in the particle domain for a full time step $\Delta t$ using the Euler-Maruyama scheme of Eq. (\ref{eq:eulermaruyama}). Note the diffusion coefficient is different for different species.
		
		\item Determine and apply reactions for another half a time step, $\Delta t /2$, in the same way as in step 4.
		
		\item Inject particles for another half a time step, $\Delta t /2$, in the same way as in step 3.
		
		\item Apply boundary conditions to particles. If any particle crossed into the concentration domain, eliminate it.
	\end{enumerate}
	
	There is still some freedom in the implementation of small but relevant details in the Strang splitting. We treated the injection and reaction processes as one step in the main Strang splitting, i.e. injection/reactions for half a time step (steps 3 and 4); diffusion for a full time step (step 5); and injection/reactions for half a time step (steps 6 and 7). This implies the following rules:
	
	\begin{itemize}
		\item Particles that are injected in step 3 cannot be used for reactions in step 4. However, they can be used for reactions in step 6.  
		\item Particles that reacted in step 4 or were generated during a reaction in step 4, cannot react again in another reaction in step 4. However, they can be used for reactions in step 6.  
		\item Particles that reacted in step 6 or were generated during a reaction in step 6, cannot react again in another reaction in step 6. They can only be used for reactions in the next time step.
	\end{itemize}

   Following Eq. (\ref{eq:maxdeltat}), we further recommend to choose the time step $\Delta t$ and boundary cell width $\delta x$ to satisfy the relation $\Delta t = \delta x^2/(2D)$. This maximizes the discrete jumping probabilities \cite{del2016discrete,del2018grand}, and it is the most accurate if new particles are placed uniformly on the boundary cells of the particle domain. In general, $\Delta t \leq \delta x^2/(2D)$ must be satisfied. Note we can choose several values of $\delta x$, one per species with a different diffusion coefficient. 
	
	\section{Numerical Results}\label{sec:NumRes}
	In this section, we show the numerical results for four examples. The first example is a test case for diffusion processes to verify the coupling with a time- and spatially-dependent reservoir. The second example verifies the hybrid scheme by implementing a system with first-order reactions. The third example implements the hybrid scheme for the Lotka-Volterra system, which includes second-order reactions. This verifies the coupling for a more realistic and complex reaction system with bimolecular reactions. The last example couples a particle-based Lotka-Volterra system, where the reservoir is not modeled by a PDE but by a simple constant function. This illustrates that the coupling scheme works for cases beyond reservoirs mediated by reaction-diffusion PDEs.
	\begin{figure}[htb]
		\noindent
		\flushleft\textbf{a.} \\
		\includegraphics[width=0.95\columnwidth]{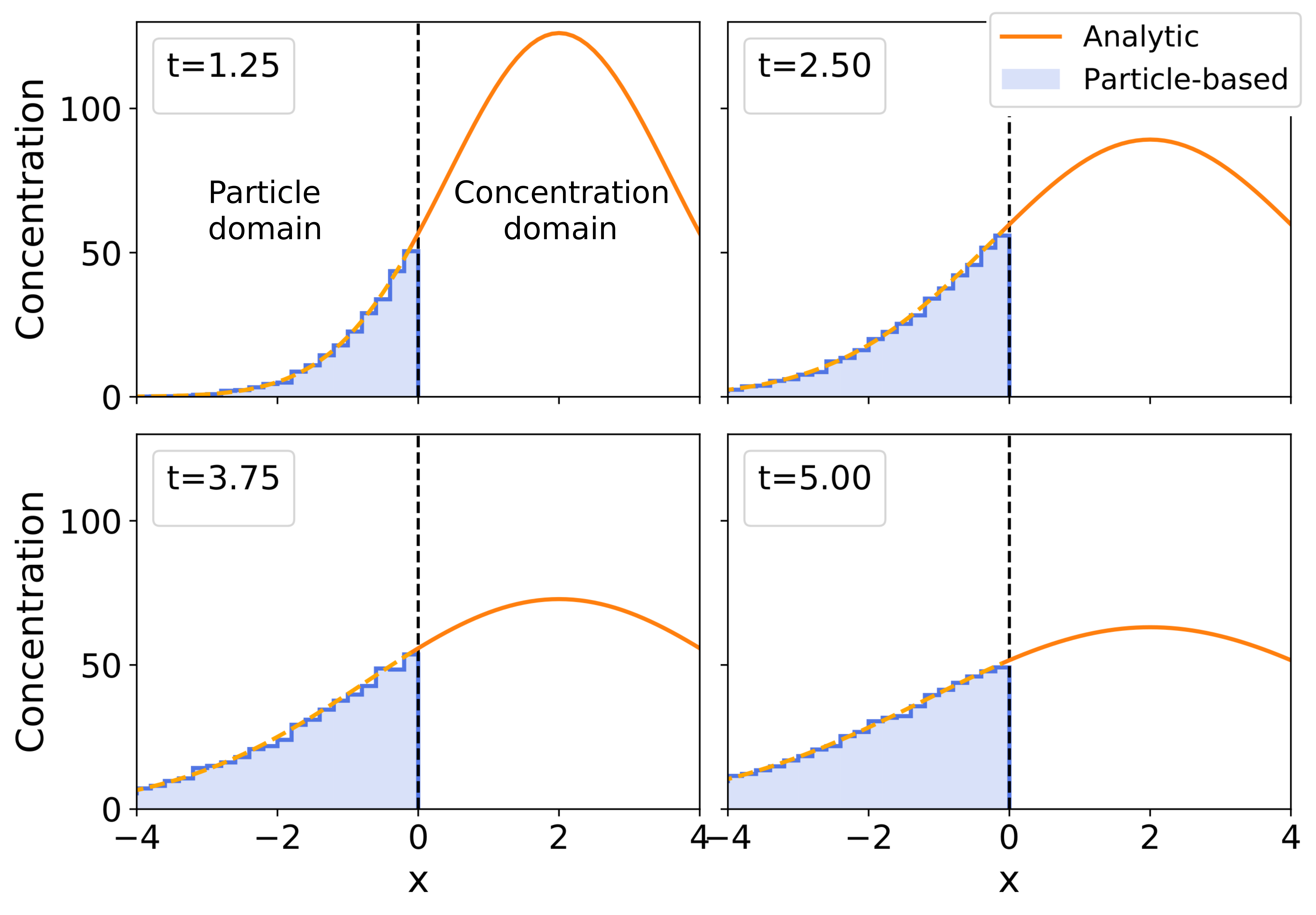} \\
		\textbf{b.} \\
		\includegraphics[width=0.95\columnwidth]{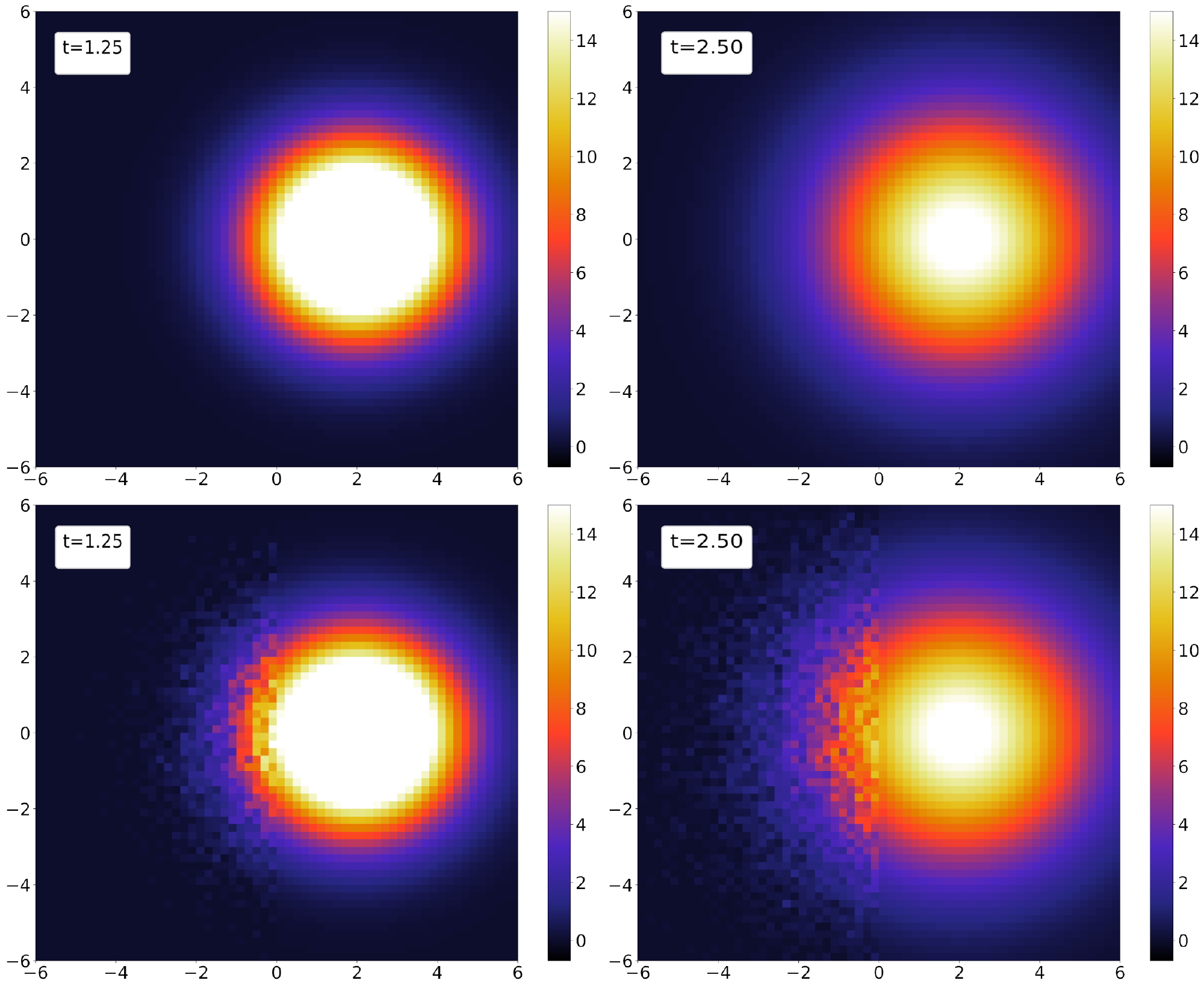} 
		\caption{Diffusion coupling results in one and two dimensions. \textbf{a.} Diffusion coupling results in one dimension at four different times. The analytic solution of the diffusion PDE with $D=1$ is shown in orange (Eq. (\ref{eq:solgauss}) with $n=1$). The average concentration of $200$ particle-based simulations in the particle domain is shown as a blue histogram. The parameters used were $\delta x = 0.05$ and $\Delta t = \delta x^2/(2D)$. \textbf{b.} Coupling results for the two dimensional extension with the same parameters, and the colorbar values denoting concentration. The solution is showed at two times. The top row shows the analytic solution (Eq. (\ref{eq:solgauss}) with $n=2$). The bottom row shows the hybrid simulation with the interface at $x=0$. In the particle domain (left half), it shows a histogram of the average concentration of $200$ particle-based simulations. In the concentration domain (right half), it shows the reservoir, which corresponds to the reference value given by the PDE.}
		\label{fig:diff1D}
	\end{figure}
	
	\begin{figure*}[htb]
		\centering 
		\includegraphics[width=\textwidth]{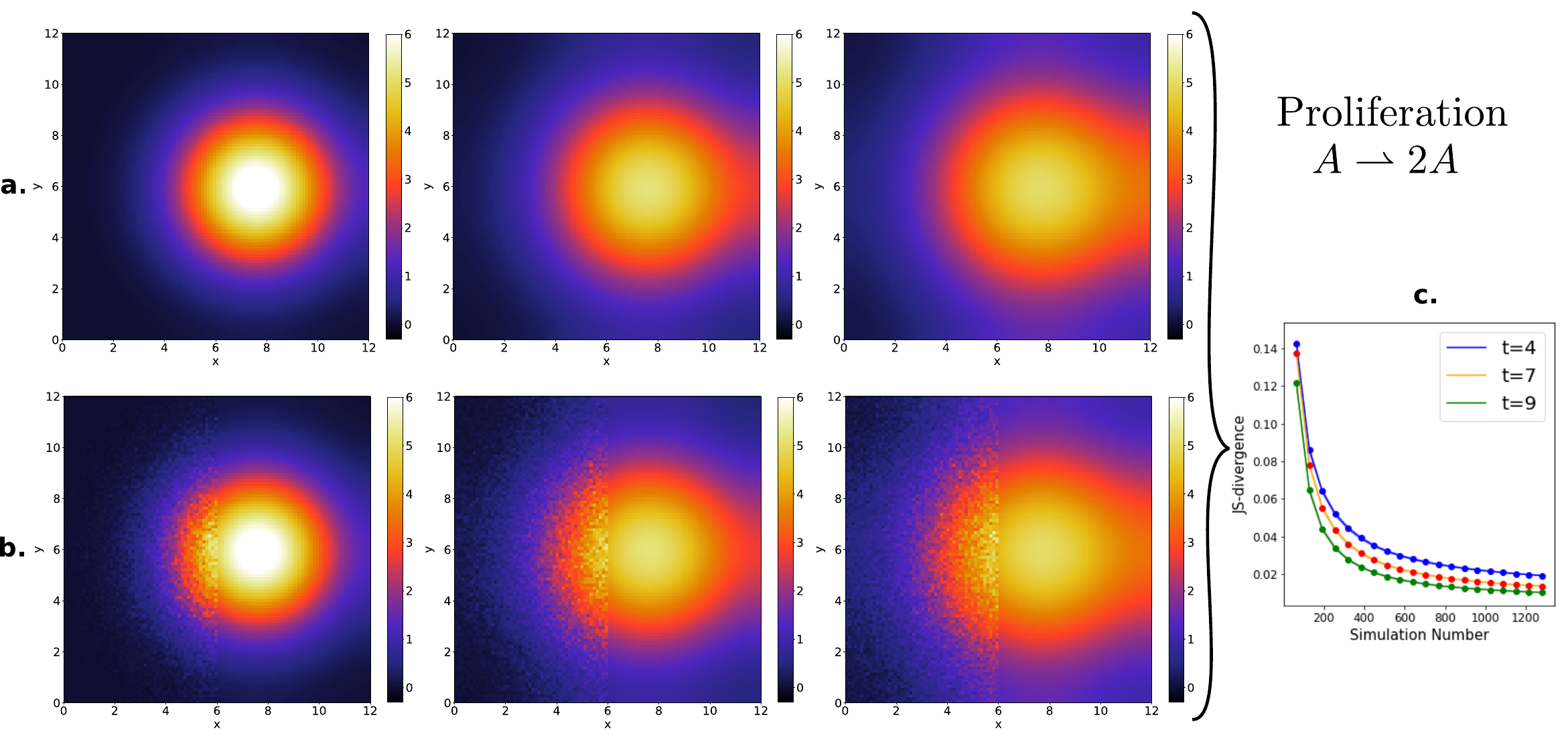}
		\caption{Coupling results for diffusion with a proliferation first-order reaction at three different times $t=4, 7, 9$. The color bar indicates the value of the concentration. \textbf{a.} Reference solution using the finite difference scheme in a domain of $12\times 12$, a grid cell size of $0.12\times0.12$ and a time step of $0.01$. \textbf{b.} Solution of the hybrid system with the interface at $x=6$. The left half corresponds to the particle domain, and it shows the average over $3000$ particle-based simulations, each using a time step of $\Delta t=0.01$. The boundary cell width $\delta x$ is chosen to satisfy $\Delta t = \delta x^2/(2D)$. The right half is the same as in the reference solution, and it serves as the material reservoir for the particle-based simulations following the scheme from Section \ref{sec:hybridscheme}. \textbf{c.} JS divergence between the reference concentration and the averaged concentration of the hybrid simulations at the same three times. The x-axis is the number of averaged hybrid simulations. Each point is calculated using $500$ bootstrapped samples.}
		\label{fig:FirstOrderResults}
	\end{figure*}
	
	\begin{figure*}[htb]
		\centering 
		\includegraphics[width=\textwidth]{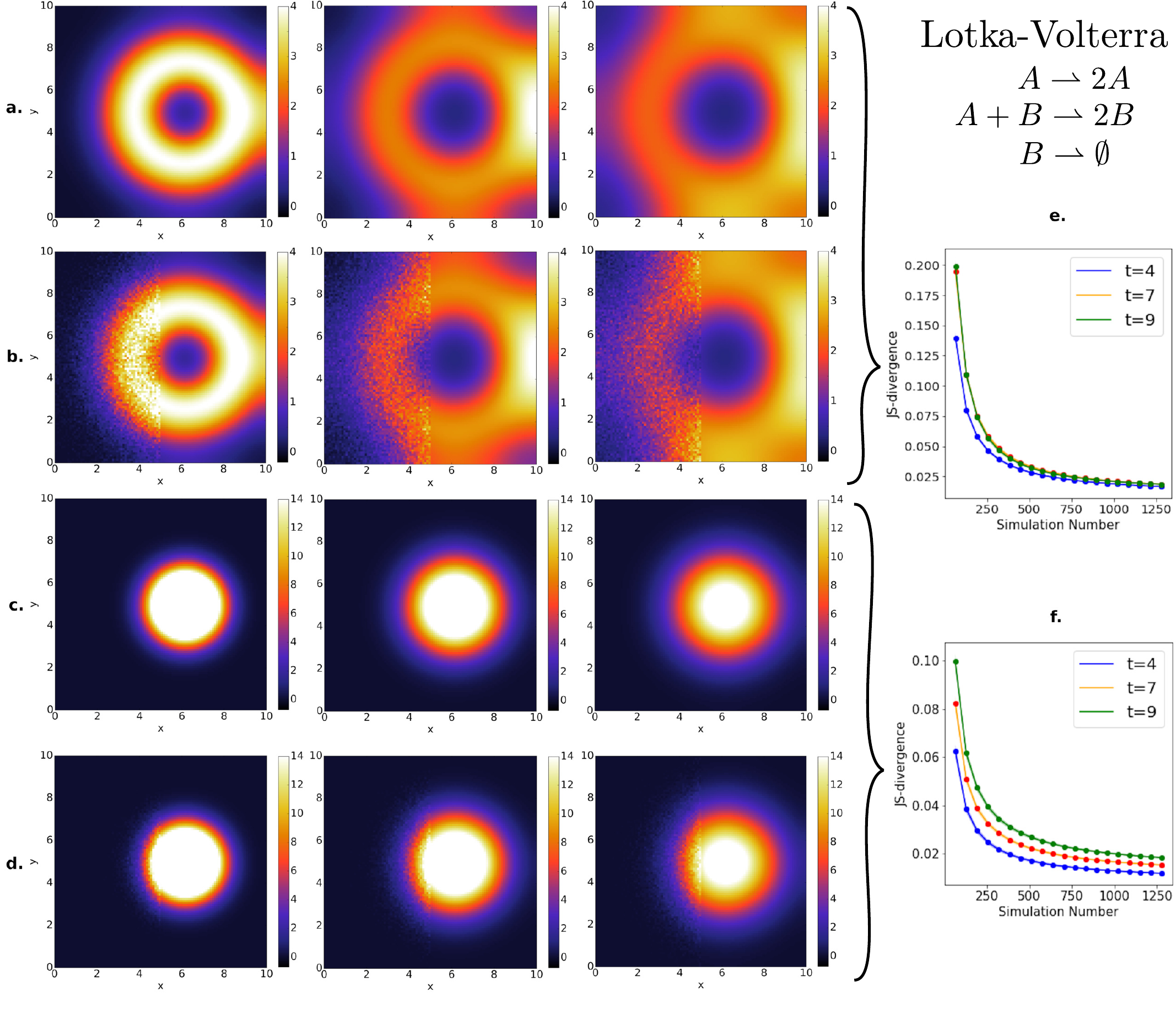}
		\caption{Solutions for the Lotka-Volterra dynamics with diffusion at three different times $t=4, 7, 9$ in a $10\times 10$ domain. The color bar indicates the value of the concentration. \textbf{a.} Reference solution of the preys using a finite difference scheme with a time step of $0.002$ and grid size of $0.1$. \textbf{b.} Solution of the preys in the hybrid simulation with the interface at $x=5$. The left half of the domain consists of the average over $3000$ particle-based simulation using a time step of $\Delta t=0.002$. We use two boundary cell widths, one for the preys ($\delta x_A$) and one for the predators ($\delta x_B$), each satisfies the relation $\Delta t = \delta x_k^2/(2D_k)$ with $k=A$ or $B$. The right half is the same as the reference solution, and it is used as the reservoir for the particle-based simulation. The coupling used is the one described in Section \ref{sec:hybridscheme}. \textbf{c.} Reference solution of the predators ($B$) corresponding to the same simulation as in Fig. \ref{fig:LVresults}a. \textbf{d.} Solution of the hybrid simulation for the predators corresponding to the same simulation as in Fig. \ref{fig:LVresults}b. \textbf{e.} \& \textbf{f.} JS divergence, for preys and predators respectively, calculated between the reference concentration and the averaged concentration of the hybrid simulations at the three times. The x-axis is the number of hybrid simulations used to calculate the average, and each point is calculated using $500$ bootstrapped samples.}
		\label{fig:LVresults}
	\end{figure*}

	\begin{figure*}[htb]
	\centering 
	\includegraphics[width=0.95\textwidth]{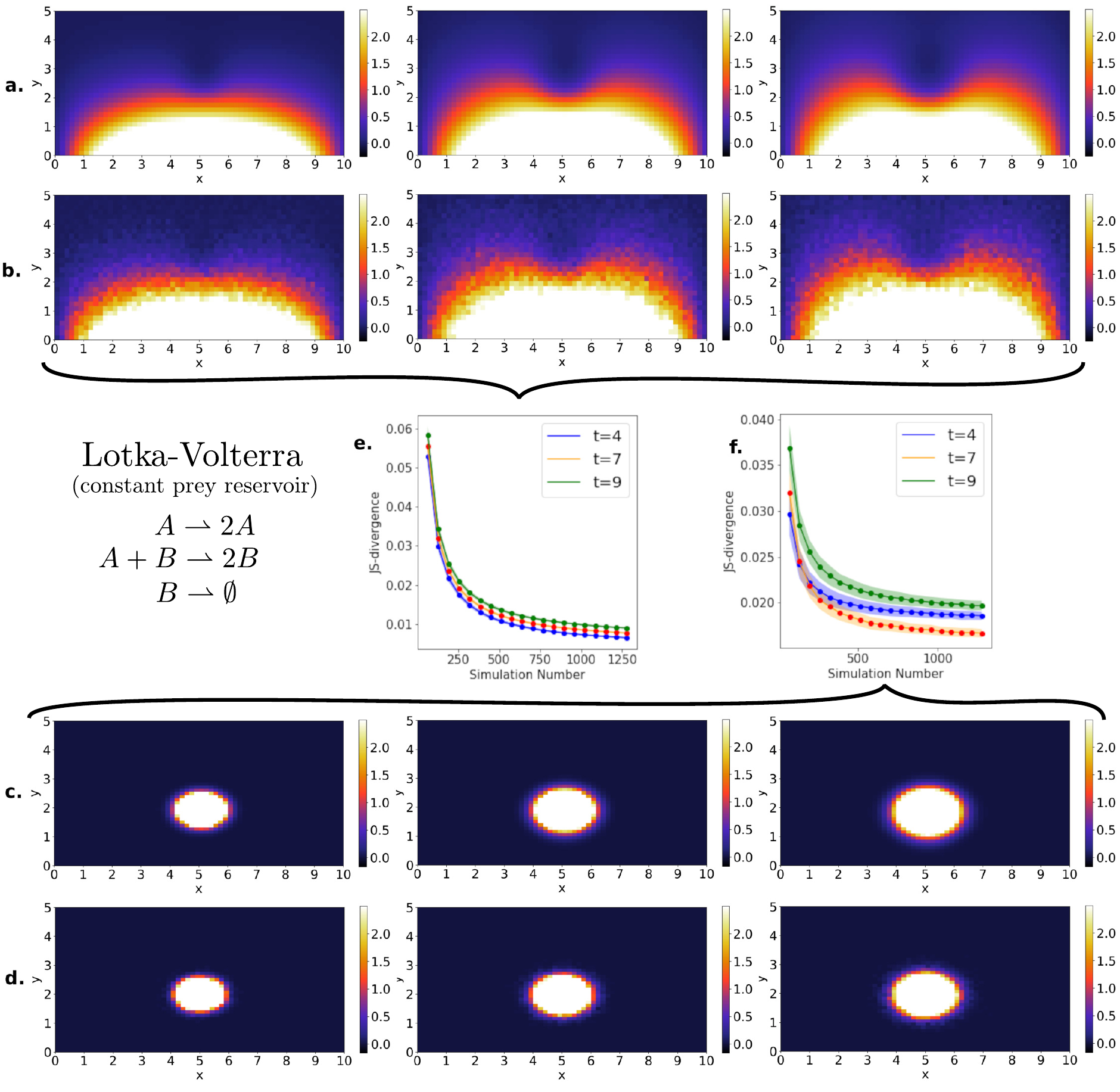}
	\caption{Coupling results for a Lotka-Volterra system coupled to a constant in time prey reservoir at three different times $t=4,7,9$. The color bar indicates the value of the concentration. \textbf{a.} Reference solution of preys using a finite difference scheme with $60\times 30$ grid cells and a time step of $0.001$. \textbf{b.} Solution of preys in the hybrid simulation, consisting of the average over $3000$ particle-based simulation. Each particle-based simulations used a time step of $\Delta t = 0.01$ and was coupled to the reservoir following the scheme from Section \ref{sec:hybridscheme}. We use two boundary cell widths, one for the preys ($\delta x_A$) and one for the predators ($\delta x_B$), each satisfies the relation $\Delta t = \delta x_k^2/(2D_k)$ with $k=A$ or $B$. \textbf{c.} Reference solution of the predators corresponding to the same simulation as in Fig. \ref{fig:sinusLV}a. \textbf{d.} Solution of the hybrid simulation for the predators corresponding to the same simulation as in Fig. \ref{fig:sinusLV}b. \textbf{e.} and \textbf{f.} JS divergence, for preys and predators respectively, calculated between the reference concentration and the averaged concentration of the hybrid simulations at the three plotted times. The x-axis is the number of averaged hybrid simulations, and each point is calculated using $500$ bootstrapped samples.}
	\label{fig:sinusLV} 
	\end{figure*}

	\subsection{Diffusion of one species}
	
	Consider the diffusion PDE with open boundaries for the concentration $c$ of a chemical species, $\partial_t c(\overline{x},t)=D\nabla^2c(\overline{x},t)$ with initial condition $c(\overline{x},0)=c_0\delta(\overline{x}-\overline{x}_0)$ and $\overline{x}_0=2,0$. The solution of this equation in $d$ dimensions is a Gaussian
	\begin{align}
	c(\overline{x},t)=c_0\frac{e^{-|\overline{x}-2|^2/4Dt} }{(4\pi D t)^{d/2}}.
	\label{eq:solgauss}
	\end{align}
	
	We apply the scheme from Section \ref{sec:hybridscheme} to this problem in one and two dimensions $d=1,2$, and show its solution in Fig. \ref{fig:diff1D}. We define the particle domain by $x\in(\infty,0]$ and the concentration domain by $x\in(0,\infty)$, and we choose $\delta x=0.05$. We use the average values of $c(\overline{x},t)$ in the boundary cell/cells delimited by $x=(0,\delta x]$ as the concentration values for the particle-based simulation reservoir at each time step. The results show a comparison between the reference analytic result and the average concentration obtained from several particle-based simulations. We can observe an excellent match, illustrating a successful coupling of a particle-based simulation to a material bath mediated by a PDE. 
	
	In the pure diffusion case, we obtain very good results even when averaging over a relatively small number of simulations. In more complex cases, we will average over a larger number of simulations, and we will verify the scheme using the Jensen-Shannon (JS) divergence. 
	
	\subsection{Proliferation of one species (first-order)}
	In this example, we model the diffusion of one species $A$, along with the first-order reaction 
	\begin{align*}
	A\xrightharpoonup[]{}2A.
	\end{align*}
	
	The corresponding reaction diffusion PDE is
	\begin{align}
	\partial_{t}c&=D\nabla^2 c+\kappa_1c, 
	\label{ProlifPDE}
	\end{align}
	with $c(\overline{x},t)$ the concentration of $A$. We use a diffusion coefficient of $D=0.5$ and a proliferation reaction rate of $\kappa_1=0.1$. In Fig. \ref{fig:FirstOrderResults}a, we solve this equation using a finite difference scheme on the domain $\overline{x}\in[0,12]\times[0,12]$ with an indicator function as initial condition and Neumann boundary conditions. The indicator function consists of a concentration of $50$ in a $2\times 2$ square centered at $(7.5,6)$ and $0$ elsewhere. This is be our reference solution. Figure \ref{fig:FirstOrderResults}b shows the solution using the hybrid scheme from Section \ref{sec:hybridscheme}. In the particle-based simulation, particles that proliferate are placed in the same location as the source particle.
	
	In Fig. \ref{fig:FirstOrderResults}c., we verify the scheme using the JS divergence, which is a measure of the difference between distributions or histograms. We compare the reference solution against averages of the hybrid simulation. As the number of hybrid simulations used to produce the average grows, the JS divergence becomes closer to zero, showing the expected convergence. 
	
	\subsection{Lotka-Volterra dynamics} \label{sec:ex1LV}
	The Lotka-Volterra dynamics are a chemical kinetics system, which can be understood in terms of preys $A$ and predators $B$. If a predator meets one prey, the prey can be eaten and the predator multiplies. Moreover, the prey can multiply and predators can die, both independently. This is an important system since it captures the complex dynamics that could appear in most relevant reaction-diffusion applications. For instance, the models used in epidemiology for the spread of infectious diseases, such as the SIR model, follow similar dynamics.
	
	The kinetics are condensed in the following reactions
	\begin{align}
	A&\xrightharpoonup[]{}2A  \nonumber\\
	A+B&\xrightharpoonup[]{}2B 	\label{eq:LVreaction} \\ 
	B&\xrightharpoonup[]{}\emptyset \nonumber
	\end{align} 
	If the system is not well-mixed and the numbers of predators and preys is large, the macroscopic dynamics can be described by the PDE
	\begin{align}
	\partial_t c_A=D_{A} \nabla^2 c_A + \kappa_1c_A - \kappa_2c_Ac_B, \nonumber \\
	\partial_t c_B=D_{B} \nabla^2 c_B + \kappa_2c_Ac_B - \kappa_3 c_B, 
	\label{LotkaVolterraPDE}
	\end{align}
	where the macroscopic rates, $\kappa_1$, $\kappa_2$ and $\kappa_3$, correspond to the proliferation rate of preys, the rate at which preys are eaten and the rate at which predators die, respectively.  
	
	The domain is $\overline{x}\in[0,10]\times[0,10]$. The initial condition for the preys is a concentration of $100$ in a $2\times 2$ square centered at $(6,5)$ and $0$ elsewhere. For the predators, it consists of a concentration of $10$ in a $1\times 1$ square centered at $(6,5)$. The preys proliferate with a rate of $\kappa_1=0.15$; the predators die with a rate of $\kappa_3=0.1$ and, if they are closer than a distance $\sigma=0.01$, they react with $\alpha=0.05$. The macroscopic bimolecular reaction rate is $\kappa_2=\alpha \pi \sigma^2$, following Eq. (\ref{eq:birateRel}). The preys diffuse with $D_A=0.3$ and the predators with $D_B=0.1$. We use Neumann (reflective) boundary conditions in all the boundaries. 
	
	In Fig. \ref{fig:LVresults}, we show the reference simulation and the averaged hybrid simulation solution results for both prey and predators. The coupling produces an excellent match. We further verify the results using the JS divergence, which shows convergence as the number of averaged simulations is increased.
	
	\subsection{Reservoirs constant in time}
	We focus again on the Lotka-Volterra dynamics described by Eq. (\ref{eq:LVreaction}). However, instead of coupling the particle-based simulation to a material reservoir mediated by a PDE, we couple it to a reservoir with a constant concentration in time (not in space). This is helpful when modeling a reservoir with a specific spatial distribution. 
	
	The domain is $\overline{x}\in[0,10]\times[0,5]$, and we use the same parameters as in Section \ref{sec:ex1LV}, except for $\sigma=0.02$, $\kappa_3=0.2$, $\kappa_2=\alpha \pi \sigma^2$ and $\Delta t=0.01$. The initial condition is zero preys and an indicator function with a concentration of predators of $30$ in a rectangle of $2\times 1$ centered at $(5,2)$. The system is in contact with a constant in time reservoir of prey in the bottom boundary, where the reservoir concentration is modeled by $7\sin(\pi x/10)$.
	
	
	Figure \ref{fig:sinusLV} shows and compares the results. We observe again an excellent match between the average of hybrid simulations and the reference solution. We verify the results with the JS divergence. Differences in the JS divergence at different times are because the number of particles changes in time, changing the convergence rate.

	\section{Discussion}
	The main goal of this work was to improve current modeling techniques for biochemical open systems, as they are extremely relevant to model life-related processes. In this paper, we contributed to this goal by developing models and numerical schemes, which are capable of consistently coupling particle-based reaction-diffusion processes with reservoirs mediated by reaction-diffusion PDEs, i.e. with time-dependent and spatially non-homogeneous reservoirs.
	
	The coupling was rendered possible by the two theoretical results from Section \ref{sec:coupling}. The first result derived the mean-field of a particle-based diffusion model in contact with a constant concentration reservoir. This resulted in a diffusion PDE with a constant concentration boundary condition and elucidated the relation between the reservoir dynamics in the two models. The second result derived the mean-field limit of reaction-diffusion processes with a bimolecular reaction. We recovered the corresponding reaction-diffusion PDE, and we obtained a precise connection between the microscopic and macroscopic parameters, specifically the bimolecular reaction rates and the reaction radius. Section \ref{sec:hybridscheme} further employs these two theoretical results to build the coupling numerical scheme, which can be used in any reaction-diffusion system with up to second-order reactions.
	
	In Section \ref{sec:NumRes}, we implement the coupling scheme for four representative examples: pure diffusion of one species, proliferation of one species, Lotka-Volterra dynamics and Lotka-Volterra dynamics coupled to a reservoir constant in time. In order to verify the scheme, we compare the average of several particle-based simulations with the theoretical mean-field given by the reaction-diffusion PDE. The difference between the averaged and the reference solution is quantified with the JS divergence. We obtain excellent results for all the examples.
	
	We finally point out the result from Section \ref{sec:RDprocMF} neglected the covariances when taking the expectation of the mixed Poisson process. This is a valid approximation if there is a very large number of particles. However, for mesoscopic scales in the number of particles, the covariances would yield a non-negligible modified mean-field behavior, which would possibly yield alternative algorithms. The authors are currently researching these topics \cite{kostreHydrolimit}.
	
	\section{Acknowledgements \label{sec:ack} }
	We gratefully acknowledge support by the Deutsche Forschungsgemeinschaft (grants SFB1114, projects C03 and A04), the Berlin Mathematics Research Center MATH+, Project AA1-6, and the European research council (ERC starting grant307494 ``pcCell'').

	\bibliographystyle{IEEEtran}
	\bibliography{refs_couplingRD}
	
	\renewcommand{\thesection}{\Alph{section}} \numberwithin{equation}{section}
	
	\appendix

	\section{Finite difference solution of the reaction-diffusion PDE}\label{sec:AppendixA}
	
	We will show how to solve numerically the Poisson equation 
	\begin{align*}
	D \nabla^2 u=f,
	\end{align*}
	with $D$ the diffusion constant, $f$ an arbitrary function and \textit{homogeneous Neumann boundary conditions}
	\begin{align*}
	\frac{du}{d\eta}=0
	\end{align*}    
	on a two dimensional finite domain (this corresponds to a reflecting boundary in a PBS) with a \textit{Finite Difference scheme} \cite{deuflhard2011numerische}.
	
	As the name already suggests we discretize the domain in cells of length and wide $h$, such that we can set up a linear system 
	\begin{align}
	Au=f
	\label{eq:appLinearsys}
	\end{align}
	by using the five-point stencil. This gives us the solution of $u$ for each cell. Matrix $A$ is the so called discrete \textit{Laplace operator} that depends on the boundary conditions. In our case, we will use homogeneous Neumann boundary conditions, so it is defined by
	\begin{align*}
	A:=-\frac{D}{h^{2}}\left( \begin{array}{rrrr}
	\ B & -2\mathbb{I} & 0 \cdots & 0 \\
	-\mathbb{I} & \ddots & -\mathbb{I} & \vdots \\
	\vdots & -\mathbb{I} & \ddots & -\mathbb{I} \\
	0 & \cdots & -2\mathbb{I} & B \\
	\end{array}\right) ,
	\end{align*} 
	where $\mathbb{I}$ is the corresponding identity matrix and
	\begin{align*}
	{B=\left( \begin{array}{rrrr}
		4& -2 & 0& \cdots \\
		-1 & \ddots & -1  \\
		\cdots & -1& \ddots & -1\\
		\cdots &0 & -2& 4 \\
		\end{array}\right)} 
	\end{align*} 
	and $h$ is the length of the grid cell. The solution is obtained by inverting the matrix $A$ in Eq. (\ref{eq:appLinearsys}). 
	
	It is possible to extend the Finite-Difference scheme to solve the time dependent Heat-equation
	\begin{align}
	\label{TimeEq}
	\partial_{t}u = D\nabla^2 u.
	\end{align}
	
	\noindent To do so, we can combine the Finite-Difference scheme with the Euler method. Let $u^{n}$ denote the solution at discrete time $t_{n}$. Then we can discretize in time the equation (\ref{TimeEq}) for every time step $n$ and small $\Delta t$  to obtain
	\begin{align}
	\label{Euler}
	u^{n+1}=u^{n}+\Delta t A u^{n} = [\mathbb{I} + \Delta t A]u^{n}.
	\end{align}
	If we use the \textit{implicit} Euler scheme, we obtain
	\begin{align*}
	u^{n+1}=u^{n}+\Delta t Au^{n+1} \\
	\Rightarrow [\mathbb{I}-\Delta t A]u^{n+1} = u^n,
	\end{align*}
	If we combine half a time step with Euler method and half a time step with the implicit Euler method, we obtain a very numerically robust method that is second order accurate in both space and time. It is called the Crank-Nicolson method \cite{jovanovic2013analysis, leveque2007finite},
	\begin{align*}
	\left[\mathbb{I}-\frac{1}{2}\Delta t A\right]u^{n+1} =
	\left[\mathbb{I}+\frac{1}{2}\Delta t A\right]u^{n} \\
	\Rightarrow         
	u^{n+1} =\left[\mathbb{I}-\frac{1}{2}\Delta t A\right]^{-1}
	\left[\mathbb{I}+\frac{1}{2}\Delta t A\right]u^{n}.
	\end{align*}
	We use this method throughout this work to solve the required PDE's. For the additional reaction terms, we couple this method to an operator splitting approach (Strang splitting). 
	
\end{document}